\documentclass{jfm}
\usepackage{graphicx}
\usepackage{epstopdf, epsfig}
\usepackage{psfrag}
\usepackage{subfigure}
\usepackage{color}
\usepackage{dcolumn}
\usepackage{bm}
\usepackage{amsmath,amssymb}
\usepackage{rotating,multirow,setspace,colordvi}
\usepackage{tikz}
\usepackage{pgfplots}
\usepackage{pgfplotstable}
\usepgfplotslibrary{groupplots}
\usepackage{pgfplots,capt-of}
\usepgfplotslibrary{fillbetween}

\usepackage{amsmath}
\newcommand{\specialnumber}[1]{%
  \def\tagform@##1{\maketag@@@{(\ignorespaces##1\unskip\@@italiccorr#1)}}%
}
\newcommand{\specialeqref}[2]{\begingroup
  \def\tagform@##1{\maketag@@@{(\ignorespaces##1\unskip\@@italiccorr#2)}}%
  \eqref{#1}\endgroup}

\shorttitle{High-Reynolds-number flow past a shear-free circular
cylinder} \shortauthor{A. Kumar, N. M. A. Rehman, P. Giri, R. K.
Shukla}

\title{An asymptotic theory for the high-Reynolds-number flow past a shear-free circular cylinder}

\author{Anuj Kumar\aff{1}\thanks{Present address: Department of Applied Mathematics,
Baskin School of Engineering, University of California, Santa Cruz, CA 95064, USA},
Nidhil M. A. Rehman\aff{1},
 Pritam Giri\aff{1}, Ratnesh K. Shukla\aff{1}%
\corresp{\email{ratnesh@iisc.ac.in}} }
\affiliation{\aff{1}Department of Mechanical Engineering, Indian
Institute of Science, Bangalore, 560012, India}

\begin{document}
\maketitle
\begin{abstract}
We present an asymptotic theory for analytical characterization of
the high-Reynolds-number incompressible flow of a Newtonian fluid
past a shear-free circular cylinder. The viscosity-induced
modifications to this flow are localized and except in the
neighborhood of the rear stagnation point, behave like a linear
perturbation of the inviscid flow. Our theory gives a highly
accurate description of these modifications by including the
contribution from the most significant viscous term in a
correctional perturbation expansion about an inviscid base state.
We derive the boundary layer equation for the flow and deduce a
similarity transformation that leads to a set of infinite,
shear-free-condition-incompatible, self-similar solutions. By
suitably combining members from this set, we construct an
all-boundary-condition-compatible solution to the boundary layer
equation. We derive the governing equation for vorticity transport
through the narrow wake region and determine its closed-form
solution. The near and far field forms of our wake solution are
desirably consistent with the boundary layer solution and the
well-known, self-similar planar wake solution, respectively. We
analyze the flow in the rear stagnation region by formulating an
elliptic partial integro-differential equation for the distortion
streamfunction that specifically accounts for the fully nonlinear
and inviscid dynamics of the viscous correctional terms. The drag
force and its atypical logarithmic dependence on Reynolds number,
deduced from our matched asymptotic analysis, are in remarkable
agreement with the high-resolution simulation results. The
logarithmic dependence gives rise to a critical Reynolds number
below which the viscous correction term, counterintuitively,
reduces the net dissipation in the flow field.
\end{abstract}

\section{Introduction}
\label{s:Introduction}

Fluid flows over sheer-free surfaces are radically distinct from
the ones over which the relative motion between the fluid and the
adjoining surface is forbidden by a no-slip condition. A
shear-free boundary offers no resistance to the motion of the
fluid along the surface thus allowing the fluid to slip perfectly
over it. This perfect-slip condition that prevails over a
shear-free surface has far-reaching consequences. Notably, the
finite slippage of the fluid over a shear-free boundary suppresses
vorticity generation from it, which in turn inhibits the boundary
layer formation and growth processes~\citep{leal1989vorticity}.
The likelihood of flow separation on a shear-free boundary is
therefore diminished
substantially~\citep{leal1989vorticity,legendre2009influence}. As
a consequence, surface stresses and hydrodynamic loads are
drastically reduced. This exceptional feature of a shear-free
surface is targeted in devising patterned superhydrophobic
surfaces that effectively reduce drag by inducing significant slip
over underwater
bodies~\citep{ou2004laminar,you2007effects,rothstein2010slip,muralidhar2011influence,lauga2011natmat,lohse2013slip}.
The reduction in vorticity generation and flow separation offers
additional advantages such as slip-enhanced
transport~\citep{Lohse2015slip,Haase2015slip,shukla2017slip} and
slip-induced flow
stabilization~\citep{legendre2009influence,muralidhar2011influence,seo2012numerical,
li2014effect, elongatedCylin2017slip,Atul2020slip} as well.
Advances in theoretical analysis and fundamental understanding of
flow past shear-free surfaces is of significant technological
importance and paramount for an effective realization of the full
range of their drag and dissipation reducing, and transport
enhancing capabilities.

In this work, we develop an analytical model for the
high-Reynolds-number flow past a shear-free circular cylinder. The
no-slip variant of this prototypical bluff body configuration has
been extensively investigated both experimentally and through
detailed simulations~\citep[e.g.][]{strouhal1878ueber,
von1911mechanismus, williamson1996vortex}. Presence of
hydrodynamic slip over the cylinder surface has been shown to
suppress flow separation and prominent unsteady flow features
including the Reynolds-number-dependent two- and three-dimensional
vortex shedding
patterns~\citep{legendre2009influence,shukla2017slip}.
Specifically, over a perfectly slipping cylindrical surface,
computational investigations have revealed formation of a
relatively weak unseparated boundary layer and an asymptotic
saturation of the maximum surface vorticity towards a Reynolds
number independent upper limit~\citep{legendre2009influence}.
Importantly, direct numerical simulations indicate a shear-free
cylindrical boundary to be very nearly dissipation (power loss)
minimizing over a relatively wide range of Reynolds
numbers~\citep{shukla2013minimum}. Our present investigation is,
in part, motivated by the need of an in depth insight into the
peculiar characteristics of the flow past general non-planar
shear-free surfaces that only an elaborate theoretical analysis
can facilitate.

Our analysis relies on an asymptotic expansion about an inviscid,
irrotational base state that follows from the potential flow
theory. This frictionless base state violates the shear-free
boundary condition over a non-planar perfectly slipping surface at
any finite Reynolds number. Crucially, the inviscid base state
suffers from the well-known D'Alembert's paradox for not only
cylindrical but any arbitrarily shaped perfectly slipping
boundary. To enforce shear-free condition on the perfectly
slipping cylinder, we introduce corrections in the form of a
series consisting of terms that diminish progressively with the
Reynolds number. For the most significant first-order correction
term in the asymptotic expansion, we derive the appropriate
governing equations that are uniquely applicable in each of the
distinct, yet interdependent, boundary layer, rear stagnation and
wake regions of the flow field. We subsequently determine the
interconnected explicit form of the most significant correction
term in these regions. Furthermore, by determining the dissipation
associated with the shear-free-condition-consistent and
D'Alembert's-paradox-resolved flow field, we show that the
second-order term in the asymptotic expansion of the drag
coefficient exhibits an atypical logarithmic dependence on the
Reynolds number.


The asymptotic approach adopted in our work belongs to the wider
class of well-established perturbation
methods~\citep{van1975perturbation,Hinch1991Book}. Perturbation
techniques have been used with remarkable success in the analysis
of a range of flows including boundary layers, wakes and
jets~\citep{batchelor2000introduction,schlichting2003boundary,leal2007advanced}.
Specifically, to examine the high Reynolds number boundary layer
characteristics over a spherical shear-free
surface,~\citet{JFM_Moore_1962} developed an axisymmetric,
asymptotic expansion about an inviscid base state that is given by
potential flow theory. In arriving at a correction to the
celebrated drag force expression
of~\citet{levich1949motion},~\citet{JFM_Moore_1962} relied on a
linearized asymptotic analysis of the rear stagnation and the wake
regions in addition to the boundary layer analysis. Extensions of
the analysis to an oblate ellipsoidal shear-free surface and a
stationary spherical interface that separates two fluids with
finite viscosity contrast have been considered
in~\citet{moore1965velocity} and~\citet{harper1968motion},
respectively. Owing to the relatively large viscosity contrast
between water and air, the frequently encountered water-air
interface is very nearly shear-free. As such, a large body of
theoretical work on flow past a shear-free spherical boundary
under diverse conditions arose out of an interest in bubble and
droplet dynamics. A comprehensive review of the early efforts on
theoretical modelling of the hydrodynamics of spherical and
slightly deformed, near-spherical bubbles in motion can be found
in~\citet{harper1972review}. Besides being of fundamental
importance, explicit expressions of the hydrodynamic forces
exerted on bubbles are particularly useful in theoretical and
computational investigations on the collective dynamics of bubble
swarms. Considerable theoretical and computational effort has
therefore gone into characterization of forces experienced by a
bubble undergoing motion in laminar and turbulent flow regimes.
For an exhaustive treatment of this topic, we refer the interested
readers to reviews by~\citet{Magnaudet2000review} and
\citet{Michaelides2003review}.

%
%
%
%
%
%


Our analysis and the analytical tractability of our approach are
facilitated by an effective linearization of the governing
equations over the boundary layer and wake regions. As in the case
of a spherical shear-free interface, this linearization and the
resulting simplifications are direct consequences of the formation
of a relatively weak boundary layer over the shear-free surface.
Despite this apparent similarity in the analysis, numerous crucial
differences do arise between axisymmetric spherical configuration
considered by~\citet{JFM_Moore_1962} and the cylindrical
configuration analyzed in our work. Notably, unlike the
axisymmetric spherical configuration, the similarity solutions to
the boundary layer equation for the cylindrical configuration turn
out to be incompatible with the shear-free boundary condition.
This incompatibility has necessitated development of techniques
which exploit the linearity of the boundary layer equation and
enable expression of the solution in the form of an infinite
series of distinct self-similar solutions. In addition, the
absence of a vortex stretching/contraction mechanism in two
dimensions results in substantial disparities between the
stagnation region of the planar cylindrical and the axisymmetric
spherical configurations. Specifically, the Reynolds number
dependence of the size of the stagnation region differs markedly
($\mbox{O}(\Rey^{-1/4})$ in the case of shear-free cylinder versus
$\mbox{O}(\Rey^{-1/6})$ for a shear-free sphere, $\Rey$ being the
Reynolds number). Moreover, the perturbations in the stagnation
region of a shear-free cylinder are comparable to the base
inviscid state so that the assumptions that form the basis of a
linearized analysis (small perturbations) are invalidated. This
complication in the analysis of the stagnation region is
altogether absent in a shear-free axisymmetric spherical
configuration. Importantly, our analysis reveals a striking
dissimilarity between the Reynolds number dependence of the drag
coefficient associated with flow past a shear-free sphere and the
one corresponding to flow past a shear-free cylinder. A
logarithmic dependence on the Reynolds number in the case of a
circular cylinder implies that above a critical Reynolds number,
the dissipation associated with a shear-free circular boundary
exceeds the one for the irrotational potential flow. In contrast,
the far simpler dependence on the Reynolds number in the case of
axisymmetric flow past a sphere implies that the dissipation
associated with a shear-free spherical boundary is always lower
than the one corresponding to the irrotational potential flow past
a sphere~\citep{JFM_Moore_1962}.

This paper is organized as follows. The setup consisting of a
uniform flow past a shear-free circular cylinder is described in
Section~\ref{s:ProblemDefinition}. In Section~\ref{s:Analysis}, a
perturbation expansion based asymptotic analysis is developed. The
analysis includes formulation of parabolic boundary layer and wake
vorticity transport equations in regions over which diffusion in
the wall-normal or transverse directions overwhelms the diffusion
along the streamwise direction. The flow in the neighborhood of
the rear stagnation region is quite distinct from the boundary
layer and wake regions and lacks a dominant direction for
diffusive or convective momentum transport. Our analysis of the
flow in the rear stagnation region relies on a nonlinear, elliptic
partial integro-differential equation that is formulated
specifically to account for its distinct dynamics and nonlocal
character. Our treatment of the flow in the vicinity of the rear
stagnation is detailed in Section~\ref{s:stagnation_region}. In
Section~\ref{s:powerlosscoeff}, an explicit expression for the
drag coefficient is deduced from the composite flow field
constructed by combining the solutions over the boundary layer,
rear stagnation and wake regions. The principal results and key
conclusions from this work are summarized in
Section~\ref{s:Conclusion}.

\section{The flow configuration}
\label{s:ProblemDefinition}

\begin{figure}
\centering
\includegraphics[angle=270,width=0.9\textwidth]{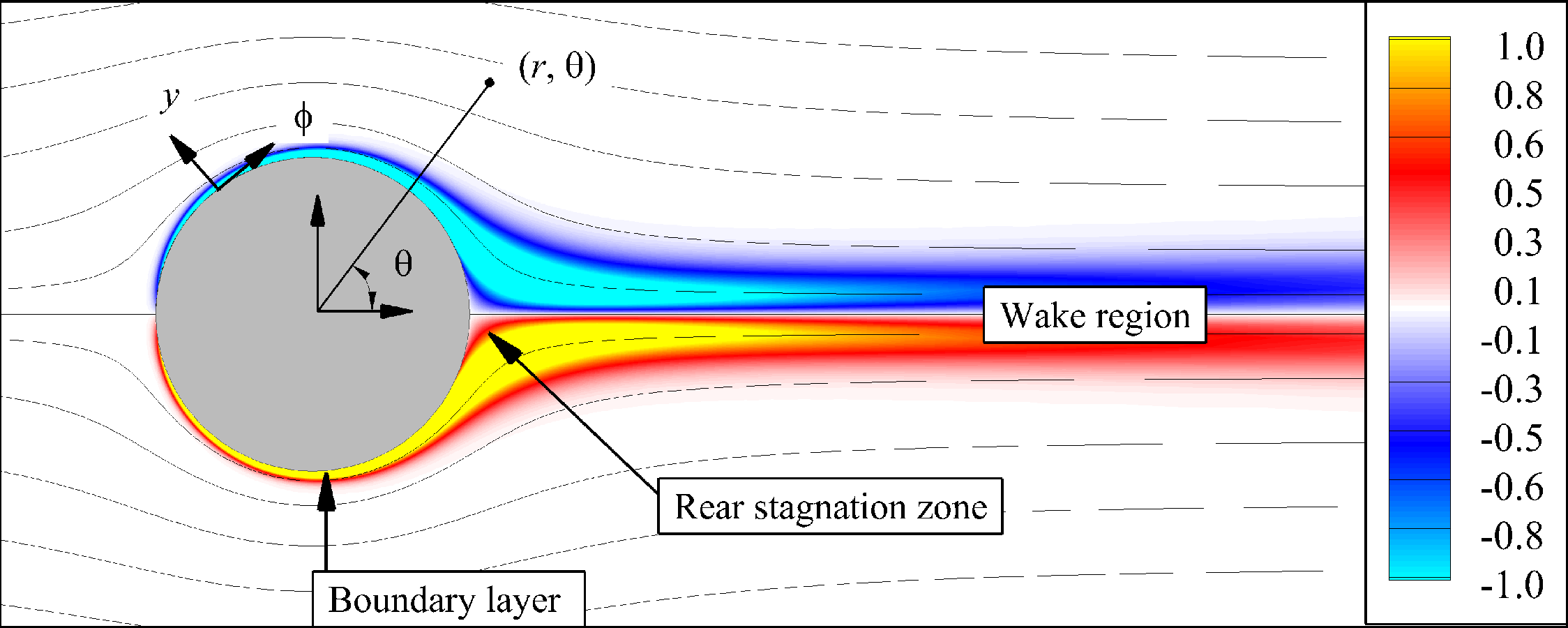}
\caption{Schematic depicting the two-dimensional flow
configuration consisting of a shear-free circular cylinder placed
in a uniform crossflow of an incompressible Newtonian fluid. The
coloured contours depict vorticity field with the overlaid thin
black lines representing the streamlines.} \label{fig:Schematic}
\end{figure}

We consider the uniform two-dimensional incompressible flow of a
viscous Newtonian fluid past a stationary, perfectly slipping
circular cylinder of diameter $D$ and infinite span. A schematic
of the setup is shown in figure~\ref{fig:Schematic}. The setup is
conveniently described in a polar coordinate system $(r,\theta)$,
with $r$ and $\theta$ denoting the radial and circumferential
coordinates, respectively. We consider steady flow, the governing
equations for which are the stationary incompressible
Navier-Stokes equations:

\begin{eqnarray}
 \rho\left( \mathbf{u}\cdot \nabla\right) \mathbf{u} = -\nabla p
 + \mu \nabla^2 \mathbf{u}, \quad \nabla \cdot \mathbf{u} =
 0,   \label{eqn:INSE}
\end{eqnarray}
where $\mathbf{u}$ and $p$ denote the velocity and pressure
fields, respectively, $\rho$ being the density and $\mu$, the
dynamic viscosity of the fluid. The following no through-flow and
shear-free boundary conditions hold on the perfectly slipping
cylinder's surface given by $r = a$ ($a = D/2$ being the radius of
the circular cylinder):
\begin{subequations}
\begin{eqnarray}
 u_r(a,\theta) &=& 0, \label{eqn:Vel-BC near-a} \\
\tau_{r\theta}(a,\theta) &=& 0 \quad \Leftrightarrow
 \frac{\partial u_{\theta}}{\partial r}\bigg|_{r=a} =
\frac{u_{\theta}(a,\theta)}{a}, \label{eqn:Vel-BC near-b}
\end{eqnarray}
\end{subequations} where $u_r$ and $u_{\theta}$ denote the radial
and circumferential components of the velocity vector
$\mathbf{u}$. In (\ref{eqn:Vel-BC near-b}), $\tau_{r\theta}$
represents the $r\theta$ component of the viscous stress tensor
$\boldsymbol{\tau} = \mu\left( \nabla \mathbf{u} + \left(\nabla
\mathbf{u}\right)^T \right)$. Sufficiently far away from the
cylinder, the flow attains a quiescent state corresponding to the
uniform free-stream along the positive $x$ direction. The
far-field boundary conditions are given by
\begin{equation}
\specialnumber{a-c} u_r \to U_\infty \cos\theta, \quad u_\theta
\to -U_\infty \sin\theta  \quad \text{and} \quad p \to p_\infty
\quad \text{as } r \to \infty, \label{eqn:Vel-BC far}
\end{equation}
where $U_\infty \mathbf{i}$ and $p_\infty$ denote the free-stream
velocity and pressure, respectively, $\mathbf{i}$ being the unit
vector in the $x$ direction. The flow is uniquely characterized by
the Reynolds number $\Rey = \rho U_\infty D/ \mu$.

\section{Asymptotic analysis}
\label{s:Analysis} Potential flow theory provides a simple means
of determining an approximation to the flow field. For the present
setup, the irrotational potential flow given by
\begin{equation}
\specialnumber{a-c} \overline{u}_r  = U_\infty \left(1  -
\frac{a^2}{r^2}\right) \cos\theta, \quad \overline{u}_\theta  =
-U_\infty  \left(1  +   \frac{a^2}{r^2}\right) \sin\theta,  \quad
\overline{p} = p_\infty + \frac{1}{2}\rho\left(U_\infty^2 -
\overline{u}_r^2 - \overline{u}_\theta^2\right),
\label{eqn:potential_flow_var}
\end{equation} is a solution of not just the incompressible Euler's equations
but also the incompressible Navier-Stokes equations
(\ref{eqn:INSE}). The term viscous potential flow is often
employed in recognition of this important characteristic of the
irrotational potential flow~\citep{Joseph2007Book}.

The potential flow solution is construed to satisfy the far-field
conditions (\ref{eqn:Vel-BC far}) and the no-through-flow
condition (\ref{eqn:Vel-BC near-a}). The corresponding tangential
shear stress on the cylinder's surface is given by
$\overline{\tau}_{r\theta}\left(a,\theta\right) = (4 \mu
U_\infty/a) \sin \theta$. Clearly, at any finite $\Rey$, the
tangential shear stress over the cylinder surface predicted from
the viscous potential flow solution does not vanish identically.
The incompatibility between the potential flow solution and the
shear-free boundary condition can also be viewed in terms of
vorticity production at the shear-free interface. At the
shear-free cylinder surface, a direct correspondence exists
between the surface vorticity and the tangential surface velocity:
$\omega(a,\theta) = 2 u_{\theta}(a,\theta)/a$, where $\omega$
denotes the vorticity component in the $z$ direction. Since
$\overline{u}_{\theta}(a,\theta)$ is finite, the foregoing
relationship would imply a contradictory existence of finite
surface vorticity in an otherwise irrotational and inviscid flow
field.

The apparent contradiction described above can be resolved by
accounting for the existence of a boundary layer in the immediate
vicinity of the shear-free cylinder surface. At high $\Rey$, this
boundary layer, like its no-slip counterpart, is thin but
significantly weaker as the relative change in the tangential
surface velocity across it is $\mbox{O}(U_{\infty}\Rey^{-1/2})$
(as opposed to $\mbox{O}(U_{\infty}\Rey^{1/2})$ for a no-slip
boundary). The vorticity produced in this thin boundary layer
region is convected downstream first through a rear stagnation
region and then eventually into a narrow wake. Modifications to
the potential flow solution in each of these regions containing
significant vorticity is necessary for elimination of an erroneous
fore-aft symmetry in the irrotational flow field given by
(\ref{eqn:potential_flow_var}). We note here that the D'Alembert's
paradox and the erroneous prediction of a
zero-net-momentum-deficit wake are both direct consequences of the
fore-aft symmetry of the potential flow solution. In the
forthcoming subsections, we develop interdependent asymptotic
analyses necessary to account for the finite vorticity production,
and its advection and diffusion into each of the distinct
subregions (boundary layer, stagnation zone and wake) of the flow
field.

\subsection{Boundary layer analysis}
\label{s:boundary_layer}

At sufficiently high Reynolds numbers ($\Rey \gg 1$), the
corrections to the potential flow solution that are necessary to
enforce the shear-free condition (\ref{eqn:Vel-BC near-b}) can be
sought within the broad purview of boundary layer theory. To
analyze the flow characteristics in the boundary layer region, we
define a scaled normal coordinate $y=(r-a)/a$ that is attached to
the cylinder surface, along with a new tangential coordinate $\phi
= \pi - \theta$. The origin of the ($y,\phi$) coordinate system
coincides with the forward stagnation point. In this newly defined
$(y,\phi)$ coordinate system, we next express the flow variables
as a superposition of the potential flow solution and a correction
over it:
\begin{equation}
\specialnumber{a-c} u_y = \overline{u}_y + \tilde{u}_y, \quad
u_\phi = \overline{u}_\phi + \tilde{u}_\phi, \quad p =
\overline{p} + \tilde{p}, \label{eqn:Split}
\end{equation}
where $\overline{u}_y$ and $\overline{u}_\phi$ are the normal and
tangential potential flow velocity components in the $(y, \phi)$
coordinate system while $\overline{p}$ denotes the pressure
deduced from the potential flow theory. The corresponding
corrections to the velocity components and the pressure field are
denoted by $\tilde{u}_y$, $\tilde{u}_\phi$ and $\tilde{p}$,
respectively. Using (\ref{eqn:Split}b) and
(\ref{eqn:potential_flow_var}b) in (\ref{eqn:Vel-BC near-b}) we
arrive at the following expression for the shear-free condition at
the surface of the cylinder $y=0$
\begin{eqnarray}
\frac{\partial \tilde{u}_{\phi}}{\partial y}\bigg|_{y = 0} =
4U_\infty\sin\phi + \tilde{u}_{\phi}|_{y = 0}.
\label{eqn:Shear_free_1}
\end{eqnarray}

Inside the boundary layer region, the thickness of the boundary
layer serves as an appropriate length scale along the normal
direction. Denoting the boundary layer thickness scaled with $a$
by $\delta$, we find that within the boundary layer
$(\partial/\partial y) \sim 1/\delta$. The potential flow velocity
components $\overline{u}_\phi$ and $\overline{u}_y$ in the
boundary layer scale as $U_{\infty}$ and $U_{\infty}\delta$,
respectively. Using the above facts and (\ref{eqn:Shear_free_1})
we deduce that in the boundary layer $\tilde{u}_\phi \sim
U_{\infty}\delta$. The divergence-free constraint on the velocity
field yields $\tilde{u}_y \sim U_{\infty}\delta^2$. To arrive at
the boundary layer equations, we define a stretched coordinate
system $(y^\ast, \phi)$ with a normalization in the normal
direction: $y^\ast = y/\delta$. Furthermore, we define the
following non-dimensional velocity components
\begin{equation}
\specialnumber{a,b}
 u_\phi^\ast = \frac{u_\phi}{U_\infty}, \qquad u_y^\ast = \frac{u_y}{U_\infty\delta}.
\label{eqn:scaled_velocities}
\end{equation}

In the stretched coordinate system $(y^\ast, \phi)$, the
circumferential momentum equation assumes the following form
\begin{eqnarray}
u_y^\ast\frac{\partial u_\phi^\ast}{\partial y^\ast} +
\frac{u_\phi^\ast}{(y^\ast \delta + 1)} \frac{\partial
u_\phi^\ast}{\partial \phi} + \frac{\delta u_y^\ast
u_\phi^\ast}{(y^\ast\delta+1)}= -\frac{1}{(y^\ast \delta + 1)\rho
U_\infty^2}\frac{\partial p}{\partial \phi} + \frac{\mu}{\rho
U_\infty a \delta^2}
\times \nonumber \\
\left(\frac{\partial^2 u_\phi^\ast}{\partial y^{\ast 2}} +
\frac{\delta}{(y^\ast\delta + 1)}\frac{\partial
u_\phi^\ast}{\partial y^\ast}- \frac{\delta^2 u_\phi^\ast}{(y^\ast
\delta + 1)^2}+ \frac{\delta^2}{(y^\ast \delta +
1)^2}\frac{\partial^2 u_\phi^\ast}{\partial \phi^{2}} + \frac{2
\delta^3}{(y^\ast \delta + 1)^2}\frac{\partial u_y^\ast}{\partial
\phi}\right). \label{eqn:theta_momentum}
\end{eqnarray}

The convective and the diffusive terms are both dominant in the
boundary layer and hence the quantity $\mu/(\rho U_\infty a
\delta^2) = 2/(\Rey\delta^2)$ should be $\mbox{O}(1)$. Without
loss of generality, we set $\delta = \sqrt{2/\Rey}$.

In the boundary layer coordinates $(y^\ast, \phi)$, the radial
momentum equation assumes the following non-dimensional form:
\begin{eqnarray}
\delta^2 u_y^\ast\frac{\partial u_y^\ast}{\partial y^\ast} +
\frac{\delta^2 u_\phi^\ast}{(y^\ast \delta + 1)}\frac{\partial
u_y^\ast}{\partial \phi} - \underline{\frac{\delta u_\phi^{\ast
2}}{(y^\ast\delta+1)}} = - \underline{ \frac{1}{\rho
U_\infty^2}\frac{\partial p}{\partial y^\ast} } +
\qquad \qquad \qquad \qquad \qquad \quad &&\nonumber \\
\delta^2\left(\frac{\partial^2 u_y^\ast}{\partial y^{\ast 2}} +
\frac{\delta}{(y^\ast \delta +1)}\frac{\partial u_y^\ast}{\partial
y^\ast} - \frac{\delta^2 u_y^\ast}{(y^\ast \delta +1)^2} +
\frac{\delta^2}{(y^\ast \delta +1)^2} \frac{\partial^2
u_y^\ast}{\partial \phi^2} - \frac{2 \delta}{(y^\ast \delta
+1)^2}\frac{\partial u_\phi^\ast}{\partial \phi}\;\right).&&
\label{eqn:radial_momentum}
\end{eqnarray}

From the above relationship, one can deduce that in the boundary
layer region, the contribution to the pressure from the inviscid
potential flow $\overline{p}$ provides the centripetal force
necessary for the flow to turn around the cylinder (signified
through the underlined terms in the foregoing relationship
(\ref{eqn:radial_momentum})). The pressure perturbation
$\tilde{p}$ therefore scales as $\mbox{O}(\delta^2)$.

Next, we invoke boundary layer approximations and retain only the
most significant terms. Using (\ref{eqn:Split}) in
(\ref{eqn:theta_momentum}) along with the potential flow variables
(\ref{eqn:potential_flow_var}a-c), to a leading order, we obtain
the following for the non-dimensional correction velocity
component along the circumferential direction $\tilde{u}_\phi^\ast
= \tilde{u}_\phi/(U_\infty\delta)$:
\begin{eqnarray}
&&\frac{\partial^2 \tilde{u}_\phi^\ast}{\partial y^{\ast 2}} = -2
y^\ast \cos\phi \frac{\partial \tilde{u}_\phi^\ast}{\partial
y^\ast} + 2\sin\phi \frac{\partial
\tilde{u}_\phi^\ast}{\partial\phi} + 2\tilde{u}_\phi^\ast
\cos\phi, \label{eqn:governing_equation_ustar}
\end{eqnarray}
With only the leading order terms retained, the boundary
conditions (\ref{eqn:Vel-BC near-b}) and (\ref{eqn:Vel-BC far})
reduce to
\begin{equation}
\specialnumber{a-c} \tilde{u}_\phi^\ast = 0 \text{ at } \phi = 0,
\quad \tilde{u}_\phi^\ast \rightarrow 0 \text{ as } y^\ast
\rightarrow\infty, \quad
\frac{\partial\tilde{u}_\phi^\ast}{\partial y^\ast} = 4\sin\phi
\text{ at } y^\ast = 0. \label{eqn:boundary_conditions_ustar}
\end{equation}
Note that the condition (\ref{eqn:boundary_conditions_ustar}a)
follows from the symmetry of the flow.

Boundary layer equations do not posses a characteristic scale and
admit a dimension reduction through similarity transformation. In
typical planar and axisymmetric flows, the reduction in dimension
facilitates simplification of the original partial differential
equation and the associated boundary conditions into an ordinary
differential equation and corresponding boundary conditions. A
similarity transformation was also employed
by~\citet{JFM_Moore_1962} in his analysis of axisymmetric boundary
layer formed over a shear-free sphere. Given its wide-ranging
success in a variety of flows lacking an inherent characteristic
scale, it is natural to seek a similarity solution of the boundary
layer equation for our present cylindrical setup.

An analysis of the boundary layer equation
(\ref{eqn:governing_equation_ustar}) allows us to establish that a
dimensional reduction of (\ref{eqn:governing_equation_ustar}) is
indeed possible, albeit for a slightly modified form of boundary
condition (\ref{eqn:boundary_conditions_ustar}c). Specifically, we
show existence of a family of similarity solutions
\begin{eqnarray}
\tilde{u}_{\phi, n}^\ast =
-\frac{(\sin\frac{\phi}{2})^{2n+2}}{\sqrt{\pi}\sin\phi}\int_0^{\pi}
\exp\left(\frac{-\eta^2}{2\cos^2\left(\frac{\alpha}{2}\right)}\right)\sin^{2n+2}\left(\frac{\alpha}{2}\right)\mbox{d}\alpha,
\end{eqnarray}
where $\eta=\sqrt2 y^\ast \cos\frac{\phi}{2}$ and $n =
0,1,2,\dots$ is a non-negative integer (see
Appendix~\ref{sec:Appendix-A} for details). Each $\tilde{u}_{\phi,
n}^\ast$ satisfies (\ref{eqn:governing_equation_ustar}),
(\ref{eqn:boundary_conditions_ustar}a,b) and the boundary
condition
\begin{equation}
\frac{\partial\tilde{u}_{\phi, n}^\ast}{\partial y^\ast}=f_n(\phi)
= \left[\sin\left(\frac{\phi}{2}\right)\right]^{2n+1} \qquad
\mbox{at} \; y^\ast=0.
\end{equation}
Unfortunately, none of these unique solutions $\tilde{u}_{\phi,
n}^\ast$ can individually be made to satisfy the boundary
condition (\ref{eqn:boundary_conditions_ustar}c). This is in
striking contrast to the case of flow past a shear-free sphere
wherein a single similarity transformation can be used to reduce
the parabolic boundary layer equation and determine its
closed-form solution~\citep{JFM_Moore_1962,leal2007advanced}.

The hindrance posed by the incompatibility between the derived
similarity solutions and the boundary condition
(\ref{eqn:boundary_conditions_ustar}c) can be overcome by
exploiting the linearity of the governing equation
(\ref{eqn:governing_equation_ustar}). As shown in the
Appendix~\ref{sec:Appendix-A}, using the relationship
\begin{equation}
\sin \phi = \sum_{n=0}^\infty w_n
\left(\sin\frac{\phi}{2}\right)^{2n+1}, \; \mbox{where} \; w_n =
2^{3-n}\prod_{i=1}^n \frac{(2i-3)}{i}.
\end{equation}
the final solution to (\ref{eqn:governing_equation_ustar}) and
(\ref{eqn:boundary_conditions_ustar}a-c), which may be expressed
as an infinite superposition of self-similar terms
($\tilde{u}_\phi^\ast = \sum_{n=0}^\infty w_n \tilde{u}_{\phi,
n}^\ast$), assumes the following form
\begin{eqnarray}
\tilde{u}_\phi^\ast\!=\!\frac{-4}{\sqrt{\pi}}\tan\left(\frac{\phi}{2}\right)\!\int_0^{\pi}\!\exp\left(\frac{-\eta^2}
{2\cos^2\left(\frac{\alpha}{2}\right)}\right)\sin^2\left(\frac{\alpha}{2}\right)
\sqrt{1 \!-
\!\sin^2\left(\frac{\alpha}{2}\right)\sin^2\left(\frac{\phi}{2}\right)}\;
\mbox{d}\alpha. \label{eqn:solution_us}
\end{eqnarray}
The above solution is in the form of superposition of infinite
self-similar terms as opposed to the more commonly encountered
solution consisting of a single self-similar term (as in the case
of a shear-free sphere~\citep{JFM_Moore_1962}).

Vorticity produced at the shear-free cylinder's surface is
convected and diffused throughout the boundary layer region. An
estimate of this vorticity $\omega_{bl}$ can be obtained from the
solution (\ref{eqn:solution_us}) as follows:
 \begin{eqnarray}
 \omega_{bl}^\ast = \frac{\omega_{bl}}{(U_\infty / a)} = \frac{a}{U_\infty}\left(\frac{\partial u_\theta}{\partial r}+ \frac{u_\theta}{r}-\frac{1}{r}\frac{\partial u_r}{\partial \theta} \right)
= -\frac{\partial \tilde{u}_\phi^\ast}{\partial y^\ast} + O(\delta) \qquad \qquad \qquad  && \nonumber \\
  \approx \!  -\sqrt{\frac{32}{\pi}} \!\sin\left(  \frac{\phi}{2}  \right) \! \int_0^{\pi} \!  \eta  \exp  \left(\frac{-\eta^2}{2\cos^2\left(\frac{\alpha}{2}\right)}\right) \!  \tan^2\left(\frac{\alpha}{2}\right) \!
 \sqrt{1 \!- \!\sin^2\left(  \frac{\alpha}{2}  \right)  \sin^2\left(  \frac{\phi}{2} \right)} \mbox{d}\alpha, &&
\label{eqn:solution_27}
\end{eqnarray} where the superscript `$^\ast$' is used to denote
non-dimensional vorticity.

\subsection{Rear stagnation region analysis}
\label{s:stagnation_region} The boundary layer analysis of the
previous subsection was based on the assumption of wall-normal
gradients being significantly larger than the stream-wise
gradients. The surface-normal flow gradients diminish
progressively with a gradual increase in the thickness of the
boundary layer along the periphery of the cylinder. In the
vicinity of the rear stagnation point ($(y,\phi) = (0,\pi)$), as
the flow undergoes a sharp turn, the reduction in flow gradients
in the surface-normal direction is so significant that they become
comparable to the ones in the circumferential direction. Thus, in
the neighborhood of the rear stagnation point, the boundary layer
assumptions do not hold and hance the analysis of the previous
section becomes unreliable. To complete the description of the
flow over the entire shear-free cylinder boundary, we next analyze
the flow in the vicinity of the rear stagnation region without
invoking boundary layer assumptions.

The flow in the neighborhood of the rear stagnation region
involves a viscous boundary layer that effuses out into an
inviscid flow (see figure~\ref{fig:Schematic} and the forthcoming
analysis). This specific flow scenario corresponds closely to the
one concerned with flow near a stagnation point on a rigid
boundary~\citep{harper1963boundary}. In particular, close to the
rear stagnation point we expect dominance of the convective terms
over the diffusive terms and an appreciable variation in vorticity
but not the velocity. To analyze the flow in the stagnation
region, we adopt the $(y, \theta)$ coordinates ($y$ as defined in
\S \ref{s:boundary_layer}) so that the origin of the coordinate
system coincides with the rear stagnation point. Our analysis
closely follows the work of~\citet{harper1963boundary}.
Specifically, a scaling analysis of the most significant terms
allows identification of two distinct regions with contrasting
scenarios. The first region in which the convective terms dominate
the viscous terms is centered around the rear stagnation point and
extends to a distance of $\mbox{O}(1)$ into the boundary layer.
The boundary layer assumptions themselves however are valid only
beyond a certain distance from the rear stagnation point. This
distance scales with the Reynolds number as
$\mbox{O}(\Rey^{-1/4})$. Therefore, there exists a region of
dimension $\mbox{O}(\Rey^{-1/4}) \ll \theta \ll \mbox{O}(1)$ over
which the boundary layer assumptions are valid and the flow itself
is inviscid.

We next develop a theoretical model for analysis of flow in the
neighborhood of the rear stagnation region. As shown below, the
analysis independently provides justification for the
aforementioned scaling arguments. In the vicinity of the
stagnation region, {\it i.e.} towards the final stage of the
boundary layer, vorticity is given by
\begin{eqnarray}
\lim_{\phi \to \pi} \omega_{bl}^\ast
=-\sqrt{\frac{32}{\pi}}\int_0^{\pi} \eta
\exp\left(\frac{-\eta^2}{2\cos^2\left(\frac{\alpha}{2}\right)}\right)
\frac{\sin^2\left(\frac{\alpha}{2}\right)}{\cos\left(\frac{\alpha}{2}\right)}\mbox{d}\alpha,
\quad \text{where} \quad \eta = \frac{y\theta}{\sqrt{2}\delta}.
\label{eqn:vorticity_stagnation_1}
\end{eqnarray}
In this final stage of the boundary layer, the velocity variations
are small. Consequently, the distortion in the streamfunction
$\tilde{\psi}$ is negligible compared to the contribution to the
stream-function from the potential flow $\overline{\psi}$. The net
stream-function $\psi$ which is the sum of the above two, assumes
the asymptotic form $\psi = \overline{\psi} + \tilde{\psi} \approx
\overline{\psi} \approx 2 U_\infty a y \theta$. Using this
asymptotic form, $\eta$ can be rewritten as follows
\begin{eqnarray}
\eta = \frac{\psi}{2 \sqrt{2} U_\infty a \delta} =
\frac{\psi_s^\ast}{2 \sqrt{2}}, \label{eqn:eta_magnitude_st}
\end{eqnarray}
where $\psi_s^\ast = \psi / (U_\infty a \delta)$ is the
non-dimensional stream-function near the rear stagnation point.
Combining expressions (\ref{eqn:eta_magnitude_st}) and
(\ref{eqn:vorticity_stagnation_1}) we obtain an expression for
vorticity in the final stage of boundary layer. Since this region
is inviscid, vorticity remains constant along streamlines. We
therefore arrive at the following expression for vorticity in the
stagnation region:
\begin{eqnarray}
\omega_s^\ast=-\frac{2}{\sqrt{\pi}}\int_0^{\pi} \psi_s^\ast
\exp\left(\frac{-\psi_s^{\ast 2}}{16
\cos^2\left(\frac{\alpha}{2}\right)}\right)\frac{\sin^2\left(\frac{\alpha}{2}\right)}
{\cos\left(\frac{\alpha}{2}\right)}\mbox{d}\alpha,
\label{eqn:vorticity_stagnation_2}
\end{eqnarray}
$\omega_s^\ast$ being the non-dimensional vorticity in the
stagnation region ($\omega_s^\ast = \omega_s a / U_\infty$).

The flow at the rear stagnation point corresponds closely to the
flow in a right-angled corner. For such a stagnation point flow,
both the directions $y$ and $\theta$ must be equally important or
in other words $y \sim \theta$. Furthermore, the flow itself must
turn around the corner so that $\psi \sim y \theta$. Furthermore,
close to the rear stagnation point $\psi \sim
\mbox{O}(\Rey^{-1/2})$ (see \ref{eqn:eta_magnitude_st}). Combining
the foregoing relationships, we deduce $y, \theta \sim
\delta^{1/2}$ or equivalently the size of the stagnation region is
$\mbox{O}(\Rey^{-1/4})$, in accordance with our assertion in the
preceding paragraphs.

In view of the above arguments, to analyze the rear stagnation
point flow, we next introduce an appropriately rescaled coordinate
system $(y_s^\ast, \theta_s^\ast)$, where $y_s^\ast =
y/\delta^{\frac{1}{2}}$ and $\theta_s^\ast =
\theta/\delta^{\frac{1}{2}}$. In our present planar setup, an
absence of vortex stretching mechanisms suggests that the
vorticity level in the stagnation zone matches that in the
boundary layer region. Thus, $\omega_s^\ast\sim O(1)$, which can
be inferred from (\ref{eqn:vorticity_stagnation_2}) as well. For a
non-dimensional vorticity that scales as $\mbox{O}(1)$, the
associated non-dimensional velocity corrections to the inviscid
flow must necessarily scale as the characteristic non-dimensional
length scale, or equivalently $\mbox{O}(\Rey^{-1/4})$.

An inspection of the expression (\ref{eqn:potential_flow_var}a,b)
reveals that in the rear stagnation region, the potential velocity
components themselves scale as $\mbox{O}(\Rey^{-1/4})$. This would
imply that both the correction and the potential flow components
exhibit similar scalings and are therefore comparable in
magnitude. Expressing vorticity as the curl of the velocity vector
and retaining only the significant terms as determined from the
appropriate scales in the rear stagnation region, we deduce
\begin{eqnarray}
 \frac{\partial \tilde{u}_{\theta_{(s)}}^\ast}{\partial y_s^\ast}-
\frac{\partial \tilde{u}_{y_{(s)}}^\ast}{\partial \theta_s^\ast} =
\omega_s^\ast, \label{eqn:stagnation_vorticity_2}
\end{eqnarray} where
$\tilde{u}_{\theta_{(s)}}^\ast = \tilde{u}_\theta / \left(
U_\infty \delta^{1/2}\right)$ and $\tilde{u}_{y_{(s)}}^\ast =
\tilde{u}_r /\left( U_\infty \delta^{1/2} \right)$ are the
non-dimensional correction velocity components in the
circumferential and radial directions, respectively. The
correction velocity components can be expressed in terms of the
distortion to the stream-function as follows
\begin{eqnarray}
\tilde{u}_{y_{(s)}}^\ast = \frac{\partial
\tilde{\psi}_s^\ast}{\partial \theta_s^\ast},\quad
\tilde{u}_{\theta_{(s)}}^\ast = -\frac{\partial
\tilde{\psi}_s^\ast}{\partial y_s^\ast}.
\label{eqn:stag_strm_vel_crrc}
\end{eqnarray}

Using equations (\ref{eqn:vorticity_stagnation_2}) and
(\ref{eqn:stag_strm_vel_crrc}) in the relationship
(\ref{eqn:stagnation_vorticity_2}), we obtain the following
governing equation for the correction velocity induced distortion
in the streamfunction
\begin{eqnarray}
 \frac{\partial^2 \tilde{\psi}_s^\ast}{\partial y_s^{\ast 2}} +
 \frac{\partial^2 \tilde{\psi}_s^\ast}{\partial \theta_s^{\ast 2}} = -\omega_s^\ast =
 \frac{2}{\sqrt{\pi}}\int_0^{\pi} (\tilde{\psi}_s^\ast + \overline{\psi}_s^\ast)
 \exp\left(\frac{-(\tilde{\psi}_s^\ast + \overline{\psi}_s^\ast)^2}{16 \cos^2\left(\frac{\alpha}{2}\right)}\right)
 \frac{\sin^2\left(\frac{\alpha}{2}\right)}{\cos\left(\frac{\alpha}{2}\right)}\mbox{d}\alpha,
\label{eqn:stag_dist_strm}
\end{eqnarray} where
$\tilde{\psi}_s^\ast = \tilde{\psi} / \left( U_\infty a
\delta\right)$ and $\overline{\psi}_s^\ast = \overline{\psi}
/\left( U_\infty a \delta\right) = 2 y_s^\ast \theta_s^\ast$ are
the non-dimensional distortion and potential flow streamfunctions
in the rear stagnation region, respectively. The correction flow
becomes unidirectional in the final stage of the boundary layer
and towards the beginning of the wake. These facts combined with
the no-through-flow condition at the cylinder surface and the
symmetry condition along the rear axis of symmetry ($\theta = 0$),
give rise to the following boundary conditions on the distortion
streamfunction $\tilde{\psi}_s^\ast$:
\begin{equation}
\specialnumber{a-c} \tilde{\psi}_s^\ast = 0\; \text{ at }
y_s^\ast, \theta_s^\ast=0 , \qquad \frac{\partial
\tilde{\psi}_s^\ast}{\partial y_s^\ast} \rightarrow 0\; \text{ as
} y_s^\ast\rightarrow\infty, \qquad \frac{\partial
\tilde{\psi}_s^\ast}{\partial \theta_s^\ast} \rightarrow 0\;
\text{ as } \theta_s^\ast\rightarrow\infty.
\label{eqn:boundary_conditions_stag}
\end{equation}

Equation (\ref{eqn:stag_dist_strm}) is an elliptic partial
integro-differential equation that involves a nonlocal source
term. The nonlinearity of (\ref{eqn:stag_dist_strm}) makes its
analytical tractability extremely challenging, if not impossible.
This complication arises principally because the perturbations to
the potential flow turn out to be comparable to the flow itself in
the rear stagnation region. A similar nonlinearity was encountered
in the planar analysis of two-dimensional stagnation point
flow~\citep{harper1963boundary}. We note here that in an
axisymmetric configuration, a reduction in the magnitude of
vorticity due to contraction of vortex-lines in the stagnation
region ensures that perturbations remain insignificant in
comparison with the base potential flow~\citep{harper1968motion}.
This insignificance of the perturbations facilitates full
analytical resolution of the axisymmetric flow in the vicinity of
the rear stagnation region of a shear-free spherical
surface~\citep{JFM_Moore_1962}.

\begin{figure}\small
\centering
\begin{minipage}[b]{.495\textwidth}
\centering
\begin{tikzpicture}
\begin{axis}[name=plot1,
ylabel style={overlay}, yticklabel style={overlay}, ymin=-30,
ymax=0, xmin=0, xmax=1, width=\linewidth,
xmajorticks=true,ymajorticks = true, ylabel={Scaled correction
$\tilde{u}_\phi^{\ast}$},xlabel={$\phi/\pi$}, legend
style={cells={anchor=west},at={(0.0,0.0)},anchor=south west}, tick
label style={font=\footnotesize}, every axis plot/.append
style={thick} ] \addplot [no marks, black] table
{BL-Analysis.dat}; \addlegendentry{{\color{black}BLA}} \addplot+
[no marks, red] table {Theory-100.dat};
\addlegendentry{{\color{red} RSRA ($\Rey = 10^2$)}} \addplot+ [no
marks, red, dashed] table {DNS-100.dat};
\addlegendentry{{\color{red}DNS ($\Rey = 10^2$)}}
\end{axis}
\end{tikzpicture}
\label{fig:Re-100}
\end{minipage}%
\begin{minipage}[b]{.495\textwidth}
\centering
\begin{tikzpicture}
\begin{axis}[name=plot2,
ylabel style={overlay}, yticklabel style={overlay}, ymin=-30,
ymax=0, xmin=0, xmax=1, width=\linewidth,
xmajorticks=true,ymajorticks = true, ylabel={Scaled correction
$\tilde{u}_\phi^{\ast}$},xlabel={$\phi/\pi$}, legend
style={cells={anchor=west},at={(0.0,0.0)},anchor=south west}, tick
label style={font=\footnotesize}, every axis plot/.append
style={thick} ] \addplot [no marks, black] table
{BL-Analysis.dat}; \addlegendentry{{\color{black} BLA}} \addplot+
[no marks, blue] table {Theory-1000.dat};
\addlegendentry{{\color{blue}RSRA ($\Rey = 10^3$)}} \addplot+ [no
marks, blue, dashed] table {DNS-1000.dat};
\addlegendentry{{\color{blue}DNS ($\Rey = 10^3$)}}
\end{axis}
\end{tikzpicture}
\label{Re-1000}
\end{minipage}\\ 
\begin{minipage}[b]{.495\textwidth}
\centering
\begin{tikzpicture}
\begin{axis}[name=plot1,
ylabel style={overlay}, yticklabel style={overlay}, ymin=-30,
ymax=0, xmin=0, xmax=1, width=\linewidth,
xmajorticks=true,ymajorticks = true, ylabel={Scaled correction
$\tilde{u}_\phi^{\ast}$},xlabel={$\phi/\pi$}, legend
style={cells={anchor=west},at={(0.0,0.0)},anchor=south west}, tick
label style={font=\footnotesize}, every axis plot/.append
style={thick} ] \addplot [no marks, black] table
{BL-Analysis.dat}; \addlegendentry{{\color{black} BLA}} \addplot+
[no marks, teal] table {Theory-10000.dat};
\addlegendentry{{\color{teal}RSRA ($\Rey = 10^4$)}} \addplot+ [no
marks, teal, dashed] table {DNS-10000.dat};
\addlegendentry{{\color{teal}DNS ($\Rey = 10^4$)}}
\end{axis}
\end{tikzpicture}
\label{fig:Re-10000}
\end{minipage}%
\begin{minipage}[b]{.495\textwidth}
\centering
\begin{tikzpicture}
\begin{axis}[name=plot2,
ylabel style={overlay}, yticklabel style={overlay}, ymin=-30,
ymax=0, xmin=0, xmax=1, width=\linewidth,
xmajorticks=true,ymajorticks = true, ylabel={Scaled correction
$\tilde{u}_\phi^{\ast}$},xlabel={$\phi/\pi$}, legend
style={cells={anchor=west},at={(0.0,0.0)},anchor=south west}, tick
label style={font=\footnotesize}, every axis plot/.append
style={thick} ] \addplot [no marks, black] table
{BL-Analysis.dat}; \addlegendentry{{\color{black}BLA}} \addplot+
[no marks, brown] table {Theory-100000.dat};
\addlegendentry{{\color{brown}RSRA ($\Rey = 10^5$)}} \addplot+ [no
marks, brown, dashed] table {DNS-100000.dat};
\addlegendentry{{\color{brown}DNS ($\Rey = 10^5$)}}
\end{axis}
\end{tikzpicture}
\label{Re-100000}
\end{minipage}
\caption{The scaled correction in the tangential surface velocity
($\tilde{u}_\phi^{\ast} = \sqrt{\Rey / 2} \left(u_\phi -
\overline{u}_\phi\right)/U_{\infty}$) along the shear-free
cylinder surface at Reynolds numbers $10^2$, $10^3$, $10^4$ and
$10^5$. Predictions from the boundary layer analysis (BLA) and the
rear stagnation region analysis (RSRA) are indicated using solid
black and solid coloured lines, respectively. Dotted coloured
lines represent results from direct numerical simulations (DNS).}
\label{fig:velocity_correction_surface}
\end{figure}


To make further progress we solve equation
(\ref{eqn:stag_dist_strm}) subject to boundary conditions
(\ref{eqn:boundary_conditions_stag}a-c) numerically using standard
approximation techniques. The details are given in the
Appendix~\ref{sec:Appendix-B}. Our analysis of the rear stagnation
and the boundary layer regions provides a complete description of
the flow along the periphery and in the immediate neighborhood of
the shear-free cylinder.
Figure~\ref{fig:velocity_correction_surface} depicts the scaled
tangential velocity correction $\tilde{u}_\phi^{\ast} = \sqrt{\Rey
/ 2} \left(\tilde{u}_\phi/U_{\infty}\right)$ along the surface of
the cylinder predicted from our analysis of the present and
preceding subsections at $\Rey = 10^2$, $10^3$, $10^4$ and $10^5$.
Coloured solid lines in
figure~\ref{fig:velocity_correction_surface} illustrate the trends
for the rear stagnation region obtained from the numerical
solution of (\ref{eqn:stag_dist_strm}) and
(\ref{eqn:boundary_conditions_stag}a-c). The numerical solution is
only valid in the neighborhood of the rear stagnation point. Hence
the coloured solid lines extend only over a limited range of
$\phi$ representing a portion of the rearward cylinder surface.
The scaled tangential velocity correction increases in magnitude
and subsequently decreases after attaining a maxima as one
traverses from the rear stagnation point to the forward stagnation
point along the cylinder surface. This broad trend is observed at
all the four Reynolds numbers, with both the magnitude of the peak
and the spread in $\tilde{u}_\phi^{\ast}$ strongly dependent on
the Reynolds number. We note here that this apparent dependence on
$\Rey$ is entirely due to the disparity between normalization
scales employed in figure~\ref{fig:velocity_correction_surface}
and the relevant characteristic scales in the stagnation region
($\delta$ versus $\sqrt{\delta}$ for the velocity scale for
instance).

Solid black lines in figure~\ref{fig:velocity_correction_surface}
depict the scaled tangential velocity correction deduced from the
boundary layer analysis ({\it i.e.} from equation
(\ref{eqn:solution_us})). Owing to the boundary layer specific
characteristic scale based normalization employed in
figure~\ref{fig:velocity_correction_surface},
$\tilde{u}_\phi^{\ast}$ given by (\ref{eqn:solution_us}) is
independent of the Reynolds number and so are the solid black
lines depicted in each of the four frames of
figure~\ref{fig:velocity_correction_surface}. The scaled
correction $\tilde{u}_\phi^{\ast}$ grows monotonically with
$\phi$. The growth is particularly pronounced over the rearward
cylinder surface ($\phi \geq \pi/2$). In particular,
$\tilde{u}_\phi^{\ast}$ becomes unbounded in the vicinity of the
rear stagnation point. This unphysical divergence of
$\tilde{u}_\phi^{\ast}$ as $\phi \rightarrow \pi$ is a direct
consequence of the breakdown of the assumptions inherent in the
boundary layer analysis.

To determine $\tilde{u}_\phi^{\ast}$ over the entire cylinder
periphery, we must employ complementary boundary layer and
stagnation region analyses over their respective domains of
validity. Furthermore, a matching procedure must be invoked over
the overlap region $\mbox{O}(\Rey^{-1/4}) \ll \theta \ll
\mbox{O}(1)$ to obtain a uniformly valid solution
$\tilde{u}_\phi^{\ast}$ over the entire range $0 \leq \phi \leq
\pi$. The dependence of $\tilde{u}_\phi^{\ast}$ on $\phi$, as
illustrated in figure~\ref{fig:velocity_correction_surface},
suggests that such a matching is indeed possible, albeit at
sufficiently high $\Rey$, which is when our asymptotic analyses of
boundary layer and stagnation regions are strictly valid ({\it
i.e.} in the limit $\Rey \rightarrow \infty$).

Our theoretical results can be directly compared with the
predictions from detailed simulations. To this end we perform
direct numerical simulations (DNS) of two-dimensional unsteady
incompressible flow past a shear-free circular cylinder using a
well established technique in polar cylindrical
coordinates~(see~\citet{shukla2005derivation}
and~\citet{shukla2007very} for details and verification tests). In
brief, our simulation methodology relies on a combination of
$10^{\mbox{th}}$-order compact finite difference and
Fourier-spectral schemes (in $r$ and $\theta$ coordinates,
respectively) and a second-order semi-implicit projection
scheme~\citep{hugues1998improved} for spatiotemporal
discretization. In all our runs we employ spatiotemporal
resolutions necessary to resolve the thin boundary layers and
enforce CFL (Courant-Friedrichs-Levy) convective stability
criterion. We also perform long-time simulations on successively
refined meshes to ensure mesh independence of the computed
solution.

The predictions from our direct numerical simulation runs are
depicted in figure~\ref{fig:velocity_correction_surface} alongside
results from the asymptotic analysis of the boundary layer and
rear stagnation regions. The comparison is quite encouraging and
more so at high Reynolds numbers. Specifically, over the windward
section of the cylinder's boundary, discernible deviations between
the DNS results and the asymptotic boundary layer analysis are
evidenced at $\Rey = 10^2$. The deviations reduce progressively
with an increase in the Reynolds number. In particular at the
highest $\Rey = 10^5$, the results from the asymptotic boundary
layer analysis and the DNS agree remarkably well all the way up to
the location at which a maxima in the magnitude of
$\tilde{u}_\phi^{\ast}$ is encountered.

We observe trends similar to the ones noted for the boundary layer
analysis in the rear stagnation region as well. Appreciable
deviations are evidenced between the predictions from the rear
stagnation region analysis and the DNS at Reynolds numbers of
$10^2$ and $10^3$. The deviations reduce considerably at $\Rey =
10^4$ and are only marginal at the highest Reynolds number of
$\Rey = 10^5$ investigated in our work. Overall, at a given
$\Rey$, the discrepancy between the results from DNS and the
stagnation region analysis is discernibly more pronounced than the
discrepancy between the DNS and boundary layer analysis
predictions. This increased discrepancy is attributable to the
more stringent restrictions on the Reynolds numbers that are
necessary in the rear stagnation region analysis ($\Rey^{-1/2} \ll
1$ for the boundary layer analysis versus $\Rey^{-1/4} \ll 1$ in
the case of rear stagnation region analysis).

\subsection{Analysis of the wake region}
\label{s:wake_analysis} To complete our description of the
vortical regions of the flow field we next develop an asymptotic
analysis of the narrow downstream wake region into which the
vorticity generated at the shear-free cylindrical surface is
eventually convected (see figure~\ref{fig:Schematic}). We adopt a
Cartesian coordinate system $(x,z)$ where both the scaled
coordinates $x$ and $z$ are normalized with respect to the
cylinder's radius $a$. The origin of the coordinate system
coincides with the rear stagnation point with the $x$ axis
pointing along the streamwise direction.

As argued in the earlier subsection, an absence of vortex
stretching mechanism in the present planar configuration implies
that the non-dimensional vorticity ($\omega^{\ast} =
a\omega/U_{\infty}$) scales as $\mbox{O}(1)$ in the boundary layer
and stagnation regions, and crucially, in the initial stages of
the wake region as well. The wake thickness is the appropriate
length scale for characterization of the wake region. Denoting the
wake thickness normalized with the cylinder radius by $\delta_w$
and making use of the non-dimensional vorticity scale of
$\mbox{O}(1)$, we deduce that the correction velocity in the
streamwise direction $\tilde{u}_x$ should scale as
$\mbox{O}(\delta_w)$. Applying the divergence-free condition on
the correction velocity field we obtain $\tilde{u}_z \sim
\mbox{O}(\delta_w^2)$ as the characteristic scale for the
component of the correction velocity in the $z$ direction.

In the wake region, the inviscid base flow velocity components
deduced from the potential flow theory can be shown to scale as
$\overline{u}_x \sim \mbox{O}(1)$ and $\overline{u}_z \sim
\mbox{O}(\delta_w)$, respectively. Balancing out the convective
and the diffusive terms in the wake, in a way similar to our
analysis of the boundary layer region, we obtain $\delta_w \sim
\delta$. Without loss of generality we set $\delta_w = \delta$.
Using the characteristic scales described above, we next define
appropriately normalized spatial coordinates and the base
potential flow and correction velocity components as follows:
\begin{equation}
\specialnumber{a-e} z^\ast = \frac{z}{\delta},\quad
\overline{u}_x^\ast = \frac{\overline{u}_x}{U_\infty},\quad
\overline{u}_z^\ast = \frac{\overline{u}_z}{U_\infty\delta},\quad
\tilde{u}_x^\ast = \frac{\tilde{u}_x}{U_\infty\delta},\quad
\tilde{u}_z^\ast = \frac{\tilde{u}_z}{U_\infty\delta^2}.
\label{eqn:scaled_wake_variables}
\end{equation}

With the above definitions in place, the equation that governs the
distribution of non-dimensional vorticity $\omega_w^\ast =
\omega_w a/U_\infty$ in the wake region assumes the following
linearized form
\begin{equation}
 \overline{u}_{x_0}^\ast \frac{\partial \omega_w^\ast}{\partial x}
 + \overline{u}_{z_0}^\ast \frac{\partial \omega_w^\ast}{\partial z^\ast}
 = \frac{\partial^2 \omega_w^\ast}{\partial z^{\ast 2}},
\label{eqn:vorticity_eqn_wake}
\end{equation}
where we have only retained the leading-order terms, and,
$\overline{u}_{x_0}^\ast$ and $\overline{u}_{z_0}^\ast$ denote the
zeroth-order terms in an expansion of $\overline{u}_x^\ast$ and
$\overline{u}_z^\ast$ about $z^\ast = 0$, respectively. Imposition
of symmetry of the flow field about $z^\ast = 0$ and the
irrotationality of the far-field flow leads to the following
boundary conditions
\begin{equation}
\specialnumber{a,b}
 \omega_w^\ast = 0  \text{ at } z^\ast = 0, \qquad\omega_w^\ast\to 0 \text{ as }z^\ast\to\infty.
\label{eqn:boundarycndtn_wake}
\end{equation}
Moreover, we require the solution $\omega_w^\ast$ of equation
(\ref{eqn:vorticity_eqn_wake}) subject to boundary conditions
(\ref{eqn:boundarycndtn_wake}) to be such that it results in a
drag-producing far wake with a constant momentum deficit. The
general solution of (\ref{eqn:vorticity_eqn_wake}) and
(\ref{eqn:boundarycndtn_wake}a,b) that satisfies the
aforementioned criterion assumes the following form
\begin{eqnarray}
\omega_w^\ast = \sum_{n}  \frac{ \lambda_n \zeta_n \exp\left(
-\frac{\zeta_n^2}{2}\right)}{ \left( x+ 1 + \frac{1}{x+ 1} +
c_n\right)}, \quad \text{where} \quad \zeta_n=\frac{z^\ast \left(1
- \frac{1}{(x+ 1)^2}\right)}
{\sqrt{2}\left(x+1+\frac{1}{x+1}+c_n\right)^{1/2}}.
\label{eqn:wake_vorticity_solution_1}
\end{eqnarray} Like the solution (\ref{eqn:solution_us}) of the boundary
layer equation (\ref{eqn:governing_equation_ustar}), the above
solution is in the form of superposition of a family of infinite
self-similar wake solutions each with strength $\lambda_n$ and
centered around the coordinates $(-(c_n+1),0)$. We require the yet
to be determined constants $\lambda_n$ and $c_n$ in the above
expression (\ref{eqn:wake_vorticity_solution_1}) to be such that
the asymptotic forms of $\omega_w^\ast$ and $\omega_{bl}^\ast$
match in the rear stagnation region. Note that since the flow in
the rear stagnation region is symmetric in $y$ and $\theta$, the
vorticity in the beginning of the wake region must match the one
associated with the final stage of the boundary layer. Matching
the general solution of the vorticity in the wake region given by
(\ref{eqn:wake_vorticity_solution_1}) and the vorticity
distribution given by (\ref{eqn:vorticity_stagnation_1}) allows us
to determine the following final form of $\omega_w^\ast$:
\begin{equation}
\omega_w^\ast = -\sqrt{\frac{2}{\pi}} \int_{-2}^{2} \frac{
\zeta_\kappa \exp\left( -\frac{\zeta_\kappa^2}{2}\right) \sqrt{4 -
\kappa^2} }{ \left( x+ 1 + \frac{1}{x+ 1} - \kappa\right)}
d\kappa, \quad \text{where} \quad  \zeta_\kappa= \frac{z^\ast
\left(1 - \frac{1}{(x+
1)^2}\right)}{\sqrt{2}\left(x+1+\frac{1}{x+1} -
\kappa\right)^{1/2}}. \label{eqn:wake_vorticity_final_expression}
\end{equation}

At a location far downstream, i.e. when $x \gg 1$,
(\ref{eqn:wake_vorticity_final_expression}) assumes the following
asymptotic form
\begin{eqnarray}
\omega_w^\ast \rightarrow -\frac{2 \sqrt{\pi}z^\ast}{x^{
3/2}}\exp\left( -\frac{z^{\ast 2}}{4 x} \right), \quad \mbox{for}
\: x \rightarrow \infty. \label{eqn:far_wake_soln}
\end{eqnarray}
The above asymptotic expression is precisely in the form of the
well-known self-similar solution for the defect velocity in the
laminar wake of a planar body with a given drag
coefficient~\citep{schlichting2003boundary}. A comparison with the
solution for the defect velocity
from~\citet{schlichting2003boundary} reveals that sufficiently far
away from the cylinder surface, the wake solution
(\ref{eqn:wake_vorticity_final_expression}) exhibits an asymptotic
decay rate that corresponds to a drag coefficient of $16\pi/\Rey$.
As shown in the forthcoming section, to a leading-order, the drag
coefficient associated with the flow past a shear-free circular
cylinder matches the aforementioned value of $16\pi/\Rey$ exactly.
Thus, our combined asymptotic analysis of the boundary layer, rear
stagnation and the wake regions is self-consistent in that the
drag predicted from the analysis, as derived in the forthcoming
section, equals the one deduced from the far-field wake signature
exactly.

\section{Viscous dissipation and drag coefficient}
\label{s:powerlosscoeff} The viscous corrections to the inviscid
base state induce a streamwise asymmetry in the otherwise
symmetric stress field. The viscous correction-induced
circumferential asymmetry in the surface stress gives rise to a
non-vanishing drag force. We next attempt to quantify the finite
drag force and in the process resolve the D'Alembert's paradox
specifically for the configuration being investigated.

An expression for the net drag force can be derived by summing up
the componentwise contributions from the pressure and viscous
stresses along the $x$ direction. Given the asymptotic form of the
solution, such an expression would be tantamount to summing up the
contributions from the zeroth-order inviscid base flow and the
first-order viscous correction to it. Owing to the symmetry of the
inviscid base flow, the contributions to the drag from the
zeroth-order terms in the asymptotic expansion for pressure and
viscous stresses must necessarily sum up to zero, thus leading to
the D'Alembert's paradox. An expression for the first-order term
in the asymptotic expansion of the drag can be readily determined
from the solutions derived in the preceding section.
Alternatively, to ensure consistency with the first-order
correction term in the wake solution, we may intuitively assert
that the first-order term in the asymptotic expansion of the drag
coefficient must necessarily equal $16 \pi/\Rey$.

Since we have determined the correctional flow field only to a
leading first-order, the aforementioned direct evaluation method
can not be used to determine any subsequent higher-order terms in
the asymptotic expansion of the drag
coefficient~\citep{kang1988drag}. Moreover, a direct evaluation of
the drag coefficient relies explicitly on the pressure
distribution over the surface of the cylinder. Determination of
first-order term in the asymptotic expansion of pressure field is
complicated by the inclusion of first-order effects in the surface
vorticity.

Notwithstanding these complications, an improved estimate of the
drag coefficient without any explicit involvement of surface
pressure distribution can nonetheless be deduced by appealing to
the overall mechanical energy balance in the flow. Specifically,
in any generic drag-producing, thrust generating or
self-propelling configuration with finite tangential surface
motion, the net energy dissipation rate must match the sum of the
rates at which work must be done to counteract the drag force and
sustain the tangential boundary motion~\citep{arakeri2013unified}.
Thus, one may express the non-dimensional total viscous
dissipation as a power loss coefficient, which for a shear-free
cylinder with vanishing tangential surface stress must equal the
drag coefficient~\citep{shukla2013minimum}:
\begin{eqnarray}
C_{PL} = \frac{\mu}{\rho U_\infty^3 a} \int_{\Omega}
\overline{\Phi}_{v} \; \mbox{d}\Omega = C_D - \frac{1}{\rho
U_\infty^3}\int_0^{2\pi} \tau_{r,\theta}(a,\theta)
u_{\theta}(a,\theta) \mbox{d}\theta = C_D,
\label{eqn:drag_dissipation_def}
\end{eqnarray} where $\Phi_v = \left(\mbox{\boldmath{$\tau$}} : \mathbf{\nabla
u}\right)/\mu $ is the dissipation function with $\Omega$ denoting
the entire flow domain excluding the cylinder's interior.
Furthermore, using $ \mbox{\boldmath{$\tau$}} : \mathbf{\nabla u}
= \mu \left( |\mbox{\boldmath{$\omega$}}|^2 + \nabla \cdot \left(
\mathbf{u} \cdot \nabla \mathbf{u} \right) \right)$ the drag
coefficient can be conveniently expressed in terms of the total
vorticity and the tangential surface velocity as
follows~\citep{Lamb1932Book,shukla2013minimum}
\begin{eqnarray}
C_{D} = C_{PL} = \frac{\mu}{\rho U_\infty^3 a}\left(
2\int_0^{2\pi} u_\theta^2(a,\theta) \mbox{d}\theta + \int_{\Omega}
\omega^2 \mbox{d}\Omega \right). \label{eqn:drag_dissipation_1}
\end{eqnarray}

A direct substitution of the irrotational potential flow solution
($\overline{u}_{\theta}(a,\theta) = -2U_{\infty}\sin \theta$ and
$\overline{\omega} = 0$) into the relationship
(\ref{eqn:drag_dissipation_1}) yields $\overline{C}_{PL} =
16\pi/\Rey$ for the power loss
coefficient~\citep{shukla2013minimum}. Thus, the power loss
coefficient deduced from the zeroth-order term in the asymptotic
expansion of the flow solution precisely equals the drag
coefficient estimated by retaining the zeroth and first-order
terms in the asymptotic expansion of the flow solution. Hence,
\begin{equation}
C_D^{(1)} = \overline{C}_{PL} = \frac{16\pi}{\Rey},
\label{eqn:Drag_Dissipation_order_1}
\end{equation} where the superscript `(1)' signifies the order to
which the terms in the asymptotic expansion of the drag
coefficient are retained. Note that in the case of a shear-free
sphere, a similar observation leads to Levich's celebrated result
$C_D^{(1)} = 48/\Rey$ for drag on a spherical
bubble~\citep{levich1949motion}. In essence, each term in the
asymptotic expansion of the power loss coefficient matches a
subsequent higher order term in the asymptotic expansion of the
drag coefficient. Hence an improved estimate for the drag
coefficient can simply be determined through a direct substitution
of the potential flow solution and the first-order correctional
terms in the relationship (\ref{eqn:drag_dissipation_1}).

We begin by estimating the contribution from the tangential
surface velocity (first term that appears in the parenthesis on
the right hand side of (\ref{eqn:drag_dissipation_1})). To a
leading order
\begin{eqnarray}
2\int_0^{2\pi} u_\theta^2(a,\theta) \mbox{d}\theta \approx 2
\int_0^{2\pi} \overline{u}_{\theta}^{2}(a,\theta) \mbox{d}\theta +
4 \int_0^{2\pi} \overline{u}_{\theta}(a,\theta)
\tilde{u}_{\theta}(a,\theta) \mbox{d}\theta.
\label{eqn:drag_dissipation_2}
\end{eqnarray} The first term on the right hand side of the above
expression (\ref{eqn:drag_dissipation_2}) corresponds to the
inviscid potential flow and therefore follows from the analysis in
the preceding paragraph. A numerical evaluation of the second term
on the right hand side of (\ref{eqn:drag_dissipation_2}) yields
\begin{eqnarray}
\frac{2 \mu}{\rho U_\infty^3 a}\int_0^{2\pi} u_\theta^2(a,\theta)
\mbox{d}\theta \approx \frac{16\pi}{\Rey} \left( 1 -
\frac{6.20}{\sqrt{\Rey}} \right). \label{eqn:drag_dissipation_3}
\end{eqnarray}

Next we consider the contribution from the total vorticity. Owing
to the symmetry of the flow, the integral term on the extreme
right of (\ref{eqn:drag_dissipation_1}) can be recast as follows
\begin{eqnarray}
 \frac{\mu}{\rho U_\infty^3 a}\int_\Omega \omega^2 d\Omega = \frac{2\mu}{\rho U_\infty^3 a}\int_{\Omega_u} \omega^2 d\Omega,
 \label{eqn:drag_dissipation_4}
\end{eqnarray}
where $\Omega_u$, with the subscript `$u$' representing upper,
denotes the region $(r,\theta)\; :\; 0 \leq \theta \leq \pi,  a
\leq r < \infty$. Substitution of a uniformly valid vorticity
field over the entire domain $\Omega_u$ would yield an estimate
for the integral (\ref{eqn:drag_dissipation_4}). For the present
setup, such an evaluation of (\ref{eqn:drag_dissipation_4}) would
amount to summing up individual contributions from the vorticity
distributions over the boundary layer, rear stagnation and wake
regions. The contributions from the boundary layer and the wake
regions can be deduced through a direct substitution of the
respective vorticity distributions (\ref{eqn:solution_27}) and
(\ref{eqn:wake_vorticity_final_expression}) into the integrand of
(\ref{eqn:drag_dissipation_4}). As detailed in
Section~\ref{s:stagnation_region}, owing to the nonlinearity of
(\ref{eqn:stag_dist_strm}), the vorticity distribution in the
stagnation region is determined numerically and therefore the
contribution to (\ref{eqn:drag_dissipation_4}) from the rear
stagnation region must be computed via numerical quadrature.

A direct quadrature-based numerical computation of the
contribution to (\ref{eqn:drag_dissipation_4}) from the rear
stagnation region is however problematic due to a logarithmic
singularity in $\psi_s$. Specifically, the singularity in
$\omega_s$ leads to a numerical prediction that diverges with mesh
refinement. The singularity is generic in that it exists in planar
stagnation point flows~\citep{harper1963boundary} and its full
resolution for the present specific configuration would presumably
require accounting for the presence of a viscous layer adjacent to
the shear-free cylindrical surface. For an estimation of the
contribution to (\ref{eqn:drag_dissipation_4}) from $\omega_s$, it
suffices to address the aforementioned lack of convergence through
a singularity subtraction technique that regularizes the
integrand.

To regularize the integrand in (\ref{eqn:drag_dissipation_4}) in
the rear stagnation region, we devise a desingularization
technique wherein we split (\ref{eqn:drag_dissipation_4}) as
follows
\begin{eqnarray}
 \frac{2\mu}{\rho U_\infty^3 a}\int_{\Omega_u} \omega^2 d\Omega =
 \underbrace{\frac{2\mu}{\rho U_\infty^3 a}\int_{\Omega_u} \omega^2_{bl, w} \; d\Omega}_{\text{I}} +
 \underbrace{\frac{2\mu}{\rho U_\infty^3 a}\int_{\Omega_s} \left(\omega_s^2 - \lim_{\theta \to 0} \omega_{bl}^2\right)
 d\Omega}_{\text{II}},
 \label{eqn:drag_dissipation_5}
\end{eqnarray} where $\Omega_{s}$ denotes the upper half of the
stagnation region. Our approach of constructing uniformly valid
asymptotic expansions ensures that in the vicinity of the rear
stagnation point, the from of singularity in the boundary layer
and wake vorticity distributions matches that of the rear
stagnation region vorticity distribution. Our desingularization
technique exploits this similarity in the nature of the
singularity to effectively eliminate it. Specifically, the
integrand of (II) is free of singularity and therefore amenable to
accurate numerical evaluation using standard quadrature rules. The
first integral (I) in the above expression
(\ref{eqn:drag_dissipation_5}) is evaluated over the entire upper
domain by making use of the analytical distributions
(\ref{eqn:solution_27}) and
(\ref{eqn:wake_vorticity_final_expression}) for vorticity in the
boundary layer and the wake regions, respectively. Furthermore, in
evaluation of (I) over the portion of $\Omega_u$ that coincides
with the rear stagnation region, we employ extensions of the
boundary layer/wake region vorticity ($\omega_{bl}$ or
$\omega_{wl}$) into the rear stagnation region and not the actual
vorticity in the rear stagnation region ($\omega_{s}$) itself.

The first integral (I) involves two regions, each with distinct
solutions; one from the boundary layer region given by
(\ref{eqn:solution_27}) and the other given by
(\ref{eqn:wake_vorticity_final_expression}) associated with the
wake region. We therefore rearrange (\ref{eqn:drag_dissipation_4})
as follows
\begin{eqnarray}
 \frac{2 \mu}{\rho U_\infty^3 a}\int_{\Omega_u} \omega^2_{bl, w} \mbox{d}\Omega =
 \frac{2\mu}{\rho U_\infty^3 a}\int_{\Omega_{bl}} \omega_{bl}^2 \;\mbox{d}\Omega
+ \frac{2\mu}{\rho U_\infty^3 a}\int_{\Omega_w} \omega_w^2
\mbox{d}\Omega. \label{eqn:drag_dissipation_6}
\end{eqnarray} To make further progress a partitioning of
the domain $\Omega_u$ into the boundary layer and wake regions
$\Omega_{bl}$ and $\Omega_w$ is necessary. Since the solutions in
the boundary layer and wake regions assume similar asymptotic
forms in the rear stagnation region, it can be reasonably asserted
that the curve $C$ dividing the two regions passes through the
rear stagnation point. Importantly, as established below, so long
as the dividing curve $C$ originates from the rear stagnation
point, the leading order terms in (I) (that we are interested in)
remain insensitive to the specifics of $C$.

In the asymptotic limit $\Rey \rightarrow \infty$, the vorticity
field undergoes a sharp decay away from the relatively thin
boundary layer, rear stagnation and wake regions. The integrands
in (\ref{eqn:drag_dissipation_6}) are therefore finite only over
these extremely thin regions with finite vorticity. We therefore
expect the dependence of (\ref{eqn:drag_dissipation_6}) on the
specifics of the dividing curve $C$ to remain localized to the
immediate neighborhood of the rear stagnation point. Consequently,
we may limit our specification of the curve $C$ to a small
neighborhood of the origin $(\theta,y) = (0,0)$. The left out
portion despite being significantly larger has no consequence on
the evaluation of (\ref{eqn:drag_dissipation_6}). Hence it
suffices to take $\theta \sim Ay^b$ to be the asymptotic form that
any general parametric representation of $C$ would assume in the
extreme vicinity of the rear stagnation point. An equivalent
representation in the wake coordinates reads $z \sim Ax^b$ and
since the wake solution holds only downstream of the rear
stagnation point, one may impose constraints $A > 0$ and $b >
1/2$.

We substitute the asymptotic form of $C$ in
(\ref{eqn:drag_dissipation_6}), while simultaneously substituting
for the boundary layer and wake solutions (\ref{eqn:solution_27})
and (\ref{eqn:wake_vorticity_final_expression}), respectively.
After some rather lengthy manipulations that are described in full
detail in the Appendix~\ref{sec:Appendix-C}, we eventually obtain
\begin{eqnarray}
&&  \frac{2 \mu}{\rho U_\infty^3 a}\int_{\Omega_u} \omega^2_{bl,
w} \mbox{d}\Omega \approx \frac{4 \sqrt{2}}{Re^{3/2}} \left(
\int_{\theta = 0}^{\pi} \int_{y^\ast = 0}^{\frac{1}{\delta}
(\frac{\theta}{A})^{1/b}} \omega_{bl}^{\ast 2} \; \mbox{d}y^\ast
\mbox{d}\theta +
\int_{x = 0}^{\infty} \int_{z^\ast = 0}^{\frac{A x^{b}}{\delta}} \omega_{w}^{\ast 2} \; \mbox{d}z^\ast \mbox{d}x \right) \nonumber \\
&& \approx  \frac{63.68 + 73.56b - 23.69 \ln A - 23.69 b \ln
\delta }{(b+1) \Rey^{3/2}} +
\frac{160.55 + 150.67b + 23.69 \ln A - 23.69  \ln \delta}{(b+1) \Rey^{3/2}} \nonumber \\
&&= \frac{216.02 + 11.85 \ln \Rey}{\Rey^{3/2}},
\label{eqn:drag_dissipation_7}
\end{eqnarray} where only the leading order terms have been
retained. Note that above result is independent of the constants
$A$ and $b$ and therefore the shape of the dividing curve $C$, as
asserted previously. This independence is not surprising since in
our estimation of (\ref{eqn:drag_dissipation_6}), we essentially
construct a single uniformly valid vorticity field from the
boundary layer and the wake region solutions. The independence is
consistent with the general expectation from the procedure of
obtaining solutions using matched asymptotic
expansions~\citep{van1975perturbation}. Specifically, the leading
order term associated with any quantity derivable from the matched
asymptotic expansion is independent of the way in which solutions
in distinct regions that share common regions of validity are
combined to form a composite solution.

Next we evaluate the integral (II). Recasting (II) in terms of the
non-dimensional quantities we obtain
\begin{eqnarray}
\frac{2\mu}{\rho U_\infty^3 a}\int_{\Omega_s} \left(\omega_s^2 -
\lim_{\theta \to 0} \omega_{bl}^2\right) \mbox{d}\Omega = \frac{4
\sqrt{2}}{Re^{3/2}} \int_{\theta_s^\ast = 0}^{\infty}
\int_{y_s^\ast = 0}^{\infty} \left( \omega_s^{\ast 2} -
\lim_{\theta \to 0} \omega_{bl}^{\ast 2} \right) \;
\mbox{d}y_s^\ast \mbox{d}\theta_s^\ast.
\label{eqn:drag_dissipation_8}
\end{eqnarray} Numerical evaluation of the above integral yields
\begin{eqnarray}
\frac{2\mu}{\rho U_\infty^3 a}\int_{\Omega_s} \left(\omega_s^2 -
\lim_{\theta \to 0} \omega_{bl}^2\right) \mbox{d}\Omega \approx
\frac{5.45}{\Rey^{3/2}}. \label{eqn:drag_dissipation_9}
\end{eqnarray} Combining the relationships (\ref{eqn:drag_dissipation_3}),
(\ref{eqn:drag_dissipation_7}) and (\ref{eqn:drag_dissipation_9})
we arrive at the following expression for the drag coefficient:
\begin{eqnarray}
C_D^{(2)} = \frac{16 \pi}{\Rey} \left(1 + \frac{0.24 \ln \Rey -
1.79}{\sqrt{\Rey}}\right), \label{eqn:drag_coefficient}
\end{eqnarray} with the superscript `(2)' indicating the order to
which the terms in the asymptotic expansion are retained. For
comparison purposes, we define a scaled correction term
$\tilde{C}_D$ by eliminating the contribution from the first-order
term in the asymptotic expansion of the drag coefficient
$C_D^{(1)}$ (or equivalently the power loss coefficient associated
with the inviscid potential flow) and subsequently normalizing
with respect to $\Rey^{-3/2}$ as shown below
\begin{eqnarray}
\tilde{C}_D = \Rey^{3/2} \left(C_D - C_D^{(1)}\right).
\label{eqn:drag_coeff_crrc_def}
\end{eqnarray}
The scaled correction term $\tilde{C}_D$ quantifies the
contributions to the drag coefficient from the second and any
higher-order terms in the asymptotic expansion of the viscous
corrections to the potential flow velocity field. Combining
(\ref{eqn:drag_coefficient}) and (\ref{eqn:drag_coeff_crrc_def})
we deduce
\begin{eqnarray}
\tilde{C}_D^{(2)} = 11.85 \ln \Rey - 90.17,
\label{eqn:drag_coeff_crrc_exp}
\end{eqnarray} for the scaled correction corresponding to the
second-order estimate of the drag coefficient
(\ref{eqn:drag_coefficient}).


\begin{figure}
\centering
\begin{tikzpicture}
\begin{loglogaxis}[
height = 7 cm, width=12cm, xlabel={$\Rey$}, ylabel={$C_D$},legend
style={at={(1.0,1.0)},anchor=north east}, legend style={legend
cell align=left}, xmin=10, xmax=2*10^5, ymin=10^-4, ymax=10, every
axis plot/.append style={very thick},axis on top] \addplot
[blue,domain=1:2*10^5,dashed,samples=201, no marks] {16*pi/x};
\addlegendentry{{\color{blue}$C_D^{(1)}$}} \addplot+
[teal,domain=1:2*10^5,samples=201,no marks] {16*pi/x +
16*pi*(0.24*ln(x)-1.79)/x^1.5};
\addlegendentry{\color{teal}{$C_D^{(2)}$}} \addplot+ [only marks,
mark=o,mark size = 2pt, red, draw=red] coordinates {(50,
0.712055027) (60, 0.621778011) (70, 0.552576005) (80, 0.49761501)
(90, 0.452811003) (100, 0.415533006) (200, 0.228603005) (300,
0.157769993) (400, 0.120433003) (500, 0.0973747) (600,
0.0817198977) (700, 0.0703973025) (800, 0.0618276) (900,
0.0551160015) (1000, 0.049718) (2000, 0.0251042005) (3000,
0.0167854) (4000, 0.0126055004) (5000, 0.0100915004) (6000,
0.00841352995) (7000, 0.00721340999) (8000, 0.00631234981) (9000,
0.00561155984) (10000, 0.00505072996) (20000, 0.00252533006)
(50000, 0.00100941001) (100000, 0.000504455005)};
\addlegendentry{{\color{red}$C_D^{DNS}$}}
\end{loglogaxis}
\end{tikzpicture}
\caption{Drag coefficient $C_D$ as a function of the Reynolds
number. Dashed lines labelled $C_D^{(1)}$ and solid lines labelled
$C_D^{(2)}$ represent theoretical estimates given by
(\ref{eqn:Drag_Dissipation_order_1}) and
(\ref{eqn:drag_coefficient}), respectively. Open circles labelled
$C_D^{DNS}$ indicate results obtained from direct numerical
simulations.} \label{fig:drag_coefficient}
\end{figure}


Our asymptotic estimates can be directly compared with the
predictions from high-resolution direct numerical simulations.
Figure~\ref{fig:drag_coefficient} depicts a comparison of the
theoretical result (\ref{eqn:drag_coefficient}) for the drag
coefficient with the predictions from DNS over more than four
decades of variation in the Reynolds number. The theoretical
result (\ref{eqn:drag_coefficient}) is in excellent agreement with
the DNS predictions, the agreement being near perfect over the
range $\Rey \gtrapprox 500$. With progressive increase in the
Reynolds number, the $\mbox{O}(\Rey^{-3/2})$ term in
(\ref{eqn:drag_coefficient}) diminishes at a rate that is higher
than the leading order $\mbox{O}(\Rey^{-1})$ term. The influence
of inclusion of $\mbox{O}(\Rey^{-3/2})$ term in the theoretical
estimate for drag coefficient is therefore expected to be most
prominent over the low Reynolds numbers range. This is clearly
evident from the trends depicted in
figure~\ref{fig:drag_coefficient}. Specifically, compared to
(\ref{eqn:Drag_Dissipation_order_1}), the asymptotic result
(\ref{eqn:drag_coefficient}) for $C_D^{(2)}$ is in appreciably
better agreement with the DNS predictions over the range $\Rey
\lessapprox 100$. Beyond $\Rey \approx 500$, the asymptotic
estimates (\ref{eqn:drag_coefficient}) and
(\ref{eqn:Drag_Dissipation_order_1}) are indistinguishable and
crucially, in remarkable agreement with the DNS predictions.

\begin{figure}
\centering
\begin{tikzpicture}
\begin{semilogxaxis}[
height = 7 cm, width=12cm, xlabel={$\Rey$}, ylabel={$\tilde{C}_D =
\Rey^{3/2}\left(C_D - 16\pi/\Rey \right)$}, legend
style={at={(1.0,0.0)},anchor=south east}, legend style={legend
cell align=left}, xmin=10, xmax=2*10^5, ymin=-120, ymax=60, every
axis plot/.append style={very thick},axis on top,extra x
ticks={2016.83},extra x tick
labels={{\color{brown}$\boldsymbol{\Rey_c}$}},extra x tick
style={grid=major, tick label style={anchor=north}}] \addplot
[blue,domain=1:2*10^5,dashed,samples=201, no marks]
{0.0};\addlegendentry{{\color{blue}$\tilde{C}_D^{(1)}$}} \addplot+
[teal,domain=1:2*10^5,samples=201,no marks] { 11.85*ln(x)-90.17};
\addlegendentry{{\color{teal}$\tilde{C}_D^{(2)}$}}
\addplot+ [only marks, mark=o,mark size = 2pt, red, draw=red]
coordinates {(50, -103.681) (60, -100.377998) (70,  -96.9281998)
(80, -93.5240021) (90, -90.2434006) (100, -87.1222) (200,
-64.2730026) (300, -50.8283997) (400, -41.8443985) (500,
-35.2882004) (600, -30.2148991) (700, -26.1245003) (800,
-22.7236996) (900, -19.8316002) (1000, -17.3141994) (2000,
-2.55156994) (3000, 4.96301985) (4000, 9.91014004) (5000,
13.5679998) (6000, 16.7084007) (7000, 19.1074009) (8000,
20.8680992) (9000, 22.6285992) (10000, 24.1844997) (20000,
34.098999) (50000, 45.7914009) (100000, 56.9252014)};
\addlegendentry{{\color{red}$\tilde{C}_D^{DNS}$}}
\addplot+[draw=none,name path=A] coordinates {(1,-200) (1,100)} ;
\addplot+[brown,loosely dashdotted,thick,no marks,name path=B]
coordinates {(2016.83,-200) (2016.83,100)} ;
\addplot+[draw=none,name path=C] coordinates {(1000000,-200)
(1000000,100)} ; \addplot+ [lightgray!8!white] fill between [of=A
and B]; \addplot+ [yellow!6!white] fill between [of=B and C];
\node at (axis cs:7500,-80) [fill=none,brown!50!black] {$\Rey >
\Rey_c$} ; \node at (axis cs:7500,-95) [fill=none,brown!50!black]
{$\tilde{C}_D^{(2)} > 0$}; \node at (axis cs:500,-80)
[fill=none,brown!50!black] {$\Rey < \Rey_c$} ; \node at (axis
cs:500,-95) [fill=none,brown!50!black] {$\tilde{C}_D^{(2)} < 0$} ;
\end{semilogxaxis}
\end{tikzpicture}
\label{fig:Correction} \caption{The scaled correction to the drag
coefficient $\tilde{C}_D$ as a function of the Reynolds number.
Dashed lines labelled $\tilde{C}_D^{(1)}$ and solid lines labelled
$\tilde{C}_D^{(2)}$ represent scaled corrections associated with
the theoretical estimates (\ref{eqn:Drag_Dissipation_order_1}) and
(\ref{eqn:drag_coefficient}), respectively. Open circles labelled
$\tilde{C}_D^{DNS}$ indicate scaled correction for the drag
coefficient computed from direct numerical simulations. The
vertical dash dotted line indicates the critical Reynolds number
$\Rey_c$ below and above which $\tilde{C}_D^{(2)} < 0$ (shown as
light grey shaded region) and $\tilde{C}_D^{(2)} > 0$ (shown as
light yellow shaded region), respectively. }
\label{fig:drag_coefficient-2}
\end{figure}

To assess the significance of the higher order
$\mbox{O}(\Rey^{-3/2})$ term in the theoretical estimate
(\ref{eqn:drag_coefficient}), we compare the scaled correction
term (\ref{eqn:drag_coeff_crrc_def}) deduced from the analysis
with the one obtained from DNS results.
Figure~\ref{fig:drag_coefficient-2} illustrates this comparison
over the full range encompassing over four orders of magnitude
variation in the Reynolds number. By definition $\tilde{C}_D^{(1)}
= 0$ irrespective of $\Rey$, as shown using blue dotted lines in
figure~\ref{fig:drag_coefficient-2}. A comparison between the
scaled correction given by (\ref{eqn:drag_coeff_crrc_exp}) with
the one computed from DNS indicates significant deviation over the
low Reynolds number range $\Rey \lessapprox 1000$. For $\Rey
\gtrapprox 2000$, the assumptions inherent in our asymptotic
analysis ($\Rey^{-1/2} \ll 1$ and $\Rey^{-1/4} \ll 1$) are met and
the scaled correction $\tilde{C}_D^{(2)}$ exhibits remarkable
agreement with the predictions from DNS. Importantly, except over
a small range of Reynolds number centered around $\Rey \sim 2000$,
compared to $\tilde{C}_D^{(1)}$, $\tilde{C}_D^{(2)}$ is
consistently in better agreement with the scaled correction
$\tilde{C}_D^{DNS}$ estimated from the detailed simulations. Thus,
the inclusion of $\mbox{O}(\Rey^{-3/2})$ term improves the
asymptotic analysis based prediction of the drag coefficient
across the entire range of Reynolds number.

The scaled correction $\tilde{C}_D^{(2)}$ is unusual in that it
explicitly depends on a logarithmic term in the Reynolds number.
This peculiar dependence is specific to the case of a shear-free
circular cylinder and is not observed in the case of a shear-free
sphere~\citep{JFM_Moore_1962}. The dependence leads to an
inversion in the sign of the scaled correction $\tilde{C}_D^{(2)}$
at a critical Reynolds number $\Rey_c \approx 2017$. The
$\mbox{O}(\Rey^{-3/2})$ term thus reduces the drag coefficient
over the range $\Rey < \Rey_c$, while increasing it for $\Rey >
\Rey_c$. The foregoing observation is supported by the DNS results
shown in figure~\ref{fig:drag_coefficient-2}, with the simulations
indicating a higher critical Reynolds number of about 2340 for an
inversion in the sign of the contribution from the higher order
terms.

The aforementioned inversion in the sign of high order
contributions has important energetic implications. Specifically,
our results indicate that over the low Reynolds number range $\Rey
< \Rey_c$, the net frictional loss associated with the
irrotational potential flow overwhelms the net loss associated
with flow past a shear-free circular cylinder. For $\Rey >
\Rey_c$, the net frictional loss associated with the irrotational
potential flow is marginally lower than the loss associated with
the shear-free cylinder boundary, with the difference between the
two diminishing progressively with the Reynolds number. In light
of the previous findings of~\citet{shukla2013minimum}, we
therefore conclude that the tangential surface velocities
corresponding to the viscous potential flow and a shear-free
condition minimize frictional loss associated with flow past an
impermeable circular cylinder at large ($\Rey > \Rey_c$) and small
($\Rey < \Rey_c$) Reynolds numbers, respectively. In stark
contrast, Moore's asymptotic analysis~\citep{JFM_Moore_1962} for
an axisymmetric spherical configuration suggests that the net
power loss associated with the potential flow overwhelms the loss
associated with a shear-free boundary, irrespective of the
Reynolds number.

Our approach of finding the net dissipation may seem overly
complicated, especially in view of the comparatively simpler
analysis involved in the estimation of drag on a shear-free
sphere~\citep{JFM_Moore_1962}. The apparent complication however
is merely due to the disparity in the contribution from the rear
stagnation region in the planar and axisymmetric configurations.
In an axisymmetric spherical setup, the $\mbox{O}(\Rey^{-11/6})$
contribution to the dissipation from the rear stagnation region is
dwarfed by the $\mbox{O}(\Rey^{-3/2})$ contribution from the wake
and boundary layer regions. A neglect of the contribution from the
rear stagnation region and a utilization of the boundary layer and
wake region solutions over the entire flow domain are both
warranted in an analysis of the $\mbox{O}(\Rey^{-3/2})$ term in
the asymptotic expansion of the drag coefficient. Such a
simplification is not possible for a planar cylindrical
configuration as all the three wake, boundary layer and rear
stagnation regions contribute to an equal order to the net
frictional loss. In our analysis of the contribution from the rear
stagnation region, an additional complication arises from the
nonlinear coupling between the correctional and inviscid base
state flow variables, with the equality in their order of
magnitudes eliminating the possibility of a linearization based
simplification.

\section{Summary}
\label{s:Conclusion} To conclude, we developed an asymptotic
theory for the high-Reynolds-number flow past a shear-free
circular cylinder. The simplest predictive theory for this flow,
namely the classical irrotational potential flow theory, suffers
from D'Alembert's paradox as it predicts a fore-aft symmetric
stress field for which the hydrodynamic resistance experienced by
the shear-free cylinder must necessarily vanish. Attributing this
paradoxical conclusion to a violation of the perfect slip boundary
condition on the shear-free cylinder surface, we introduced
viscous correctional terms to the irrotational potential flow that
account for both the finite vorticity production over the
shear-free cylinder surface and its highly efficient convection
into the wake region. We combined interdependent perturbation
expansion based analyses in each of the distinct boundary layer,
rear stagnation and wake regions over which vorticity is first
produced and subsequently convected downstream.

We demonstrated that a neglect of the relatively insignificant
streamwise gradients in a thin region surrounding the cylinder
surface leads to a linear boundary layer equation for the most
significant leading order term in the perturbation expansion of
the viscous correction to the inviscid base flow. Using a
similarity-transformation-induced dimension reduction we derived a
single-parameter-dependent family of distinct solutions to the
characteristic-scale-deficient boundary layer equation. Quite
atypically however, none of these solutions were entirely
compatible with all the necessary boundary conditions. Exploiting
the linearity of the boundary layer equation, we developed an
unconventional all-boundary-condition-compatible solution that
instead of a single self-similar term consisted of a weighted
superposition of countably infinite members from the family of
self-similar solutions.

The solution to the boundary layer equation grows unbounded in the
vicinity of the rear stagnation region. We attributed this
unphysical behaviour to a breakdown of the principal assumption of
streamwise gradients being negligibly small compared to the
surface-normal gradients. Through a careful reexamination of the
governing equations we arrived at the characteristic scales
appropriate for the rear stagnation region. We showed that in an
$\mbox{O}(a\Rey^{-1/4})$ sized region around the rear stagnation
point, the non-dimensional velocity fields for both the inviscid
base state and the first order term in the correctional
perturbation expansion are of comparable $\mbox{O}(\Rey^{-1/4})$
magnitude and crucially, are both governed by inviscid dynamics.
We noted that the foregoing observation on the magnitude of base
state and first-order correctional terms contrasted sharply with
Moore's analysis on axisymmetric flow past a shear-free
sphere~\citep{moore1965velocity}, but, was consistent with the
analysis of~\cite{harper1963boundary} for planar stagnation point
flows. We derived an elliptic partial integro-differential
equation with a nonlocal source term for the correctional velocity
induced distortion in the streamfunction associated with the
inviscid rotational flow over the rear stagnation region.
Numerical solution of the analytically intractable nonlinear
partial integro-differential equation allowed us to determine
well-behaved, bounded correctional fields that, in an overlap
region, were consistent with the boundary layer solution.
Comparisons with direct numerical simulations showed our
theoretical predictions of the first-order correctional term in
the tangential surface velocity to be in excellent agreement with
the high-resolution computations, the agreement being particularly
outstanding at a high Reynolds number of $10^5$, which is when the
critical assumptions $\Rey^{-1/2} \ll 1$ and $\Rey^{-1/4} \ll 1$
associated with the boundary layer and rear stagnation analyses
are both satisfied.

Our analysis of the wake region centered around simplifying
assumptions similar to the ones associated with our boundary layer
analysis. We showed that the high-Reynolds-number vorticity
transport through the narrow wake region is essentially governed
by a linear parabolic advection-diffusion equation. We derived an
explicit closed-form solution that, like the boundary layer
solution, consisted of a weighted superposition of infinite
self-similar wake profiles originating at distinct locations along
the axis of symmetry. We demonstrated that sufficiently far
downstream our infinite-term wake solution desirably simplified to
the well-known asymptotic form for a planar wake associated with a
two-dimensional body, the drag coefficient for which precisely
equals the leading-order estimate of $16/\Rey$ for a shear-free
circular cylinder.

Utilizing the equivalence between the energy dissipation rate and
the power expended in overcoming the drag force, we derived
explicit expressions for the first and second-order terms in the
asymptotic expansion of the drag coefficient. Our theoretical
estimate of the drag coefficient was in outstanding agreement with
the predictions from an exhaustive set of direct numerical
simulations spanning over three decades of order of magnitude
variation in the Reynolds number. Importantly, we demonstrated
that the second-order term's atypical logarithmic dependence on
the Reynolds number, predicted from our theoretical analysis, was
in remarkable agreement with the results from high Reynolds number
($\Rey \gtrapprox 2000$) direct numerical simulations. The unusual
logarithmic dependence led to identification of a critical
Reynolds number only below which the energy dissipation rate
(power loss coefficient) for the irrotational potential flow
exceeds the dissipation rate for flow past a shear-free cylinder.

To the best of our knowledge, our asymptotic analysis provides the
first theoretical prediction and quantitative assessment of the
finite-viscosity effects in the high-Reynolds-number flow past a
shear-free circular cylinder. The viscosity-induced modifications
to the inviscid base flow predicted from our analysis are far less
pronounced than for flow past a no-slip cylindrical surface.
Nonetheless, the implications of inclusion of finite-viscosity
effects in our theoretical analysis are quite significant. Our
results and analysis pave a way towards theoretical investigations
on the spatiotemporal stability of perturbation-sensitive flow
features such as boundary layers and wakes. An understanding of
the susceptibility of these instability-prone flow features to
external perturbations and an assessment of their potential to
destabilize the free-shear-enabling mechanism itself (as in the
case of superhydrophobic surfaces for instance) will be crucial in
technological applications that rely on sustained slip over
prolonged periods. Our analysis could form the basis of
theoretical investigations of flows over asymmetric shear-free
boundaries with nonconstant curvature. The curvature-driven
increase in vorticity production could enhance the Reynolds-number
dependent contrast in viscous correctional term's contribution to
the net dissipation in such flows. This would bear important
implications with regard to the optimality of a perfect slip
condition in minimizing dissipation associated with flows past
streamlined and bluff bodies.

\section*{Acknowledgements}
The authors acknowledge support received from the NVIDIA
Corporation (hardware donation program) and Supercomputing
Education and Research Center-Indian Institute of Science (runtime
on Cray XC40). R.K.S. acknowledges support received from the
Department of Science and Technology (DSTO 1329).

\section*{Declaration of interests}
The authors declare no conflict of interest.

\appendix
\section{Solution of the governing equation for $\boldsymbol{\tilde{u}_{\phi,}^\ast}$}
\label{sec:Appendix-A} Here we outline our procedure of
determining the non-dimensional correction velocity component
along the circumferential direction. The governing boundary layer
equation (\ref{eqn:governing_equation_ustar}) admits a similarity
solution. To show this we set $\tilde{u}_{\phi}^\ast =
H(\phi)F(\eta)$, where $\eta = y^\ast / g(\phi)$. Substitution of
the above form in (\ref{eqn:governing_equation_ustar}) yields
\begin{eqnarray}
F^{\prime\prime} + 2\eta\;F^{\prime} \left( g^2\cos \phi +
g\frac{dg}{d\phi} \sin \phi \right) - 2 F\left( g^2\cos \phi + g^2
\sin \phi \frac{1}{H}\frac{\mbox{d} H}{\mbox{d}\phi} \right) =
0.\qquad \label{App:Similarity_1}
\end{eqnarray}

Self-similarity is achieved when the terms inside the brackets are
constants independent of $\phi$. Without loss of generality we set
\begin{eqnarray}
2 g^2\cos \phi +  2 g\frac{dg}{d\phi} \sin \phi = 1, \qquad 2
g^2\cos \phi + 2 g^2 \sin \phi \frac{1}{H}\frac{\mbox{d}
H}{\mbox{d}\phi} = \alpha, \label{App:Similarity_2}
\end{eqnarray} where $\alpha$ is a constant. The only solution of
(\ref{App:Similarity_2}) that is also a solution of
(\ref{eqn:governing_equation_ustar}) and simultaneously satisfies
the boundary condition (\ref{eqn:boundary_conditions_ustar}a) is
\begin{eqnarray}
g(\phi) = \frac{1}{\sqrt{2}\cos\frac{\phi}{2}}, \qquad H(\phi) =
\frac{\beta \left( \sin \frac{\phi}{2} \right)^{\alpha}}{\sin
\phi}, \label{App:Similarity_sol_1}
\end{eqnarray} where $\beta$ is a constant and $\alpha > 1$. The
above functional forms of $g(\phi)$ and $H(\phi)$ yield
\begin{equation}
\frac{\partial \tilde{u}_{\phi}^\ast}{\partial y^{\ast}} =
\sqrt{2}\: \beta \:F^{\prime}(0) \left( \sin \frac{\phi}{2}
\right)^{\alpha-1}, \qquad \mbox{at} \; y^{\ast} = 0.
\label{app:BC-derivative}
\end{equation} Clearly the above expression cannot be matched with
the boundary condition (\ref{eqn:boundary_conditions_ustar}c) for
any choice of parameter $\beta$.

The linearity of the governing equation
(\ref{eqn:governing_equation_ustar}) can be exploited to make
further progress. To this end, we set $\alpha = 2n + 2$,
$\forall\; n\geqslant 0$, $n \in \mathbb{Z}$ and $\beta =
\sqrt{2}/F_n^\prime(0)$ so that the boundary condition
(\ref{app:BC-derivative}) assumes the form
\begin{equation}
\frac{\partial \tilde{u}_{\phi}^\ast}{\partial y^{\ast}} = \left(
\sin \frac{\phi}{2} \right)^{2n+1}, \qquad \mbox{at} \; y^{\ast} =
0. \label{app:BC-derivative-2}
\end{equation}

For the above choice of parameters, (\ref{App:Similarity_1})
simplifies as follows
\begin{eqnarray}
F_n^{\prime\prime} + \eta\;F_n^{\prime} - (2n+2)F_n = 0,
\label{App:governing_equation_Fn}
\end{eqnarray} where a subscript $n$ is used to denote the
$n$-specific solution. Likewise $\tilde{u}_{\phi,n}^\ast =
H_n(\phi)F_n(\eta)$ and the boundary condition
(\ref{eqn:boundary_conditions_ustar}b) transforms into
\begin{eqnarray}
F_n(\eta) \to 0 \text{ as } \eta \to \infty.
\label{App:boundary_condition_Fn}
\end{eqnarray}

The solution of (\ref{App:governing_equation_Fn}) that satisfies
the boundary condition (\ref{App:boundary_condition_Fn}) is
readily found
\begin{eqnarray}
F_n(\eta) = \int_0^{\pi}\frac{1}{\sqrt{2\pi}}
\exp\left(-\frac{\eta^2}{2\cos^2\left(\frac{\alpha}{2}\right)}\right)\sin^{2n+2}
\left(\frac{\alpha}{2}\right)\mbox{d}\alpha \quad \forall\;
n\geqslant -1 \text{ and } n \in \mathbb{Z}.
\label{App:Similarity_sol_Fn}
\end{eqnarray} To establish that (\ref{App:Similarity_sol_Fn}) is indeed a solution of
(\ref{App:Similarity_1}) we first consider the specific case of $n
= -1$:
\begin{eqnarray}
F_{-1}(\eta) &&=
\int_0^{\pi}\frac{1}{\sqrt{2\pi}}\exp\left(-\frac{\eta^2}{2\cos^2\left(\frac{\alpha}{2}\right)}
\right)\mbox{d}\alpha \nonumber \\
&&= \int_\eta^{\infty} \exp\left(-\frac{s^2}{2}\right)
\left(\int_{0}^{\pi}\frac{s}{\sqrt{2\pi}
\cos^2{\left(\frac{\alpha}{2}\right)}} \exp\left(\frac{-s^2
\tan^2\left(\frac{\alpha}{2}\right)}{2}\right)
\mbox{d} \alpha \right) \mbox{d} s \nonumber \\
&& = \int_\eta^{\infty} \exp\left(-\frac{s^2}{2}\right)
\left.\text{erf}\left(\frac{s \tan \left(\frac{\alpha}{2}\right)}
{\sqrt{2}}\right)\right|_{\alpha=0}^{\pi} \; \mbox{d}s\nonumber \\
&&= \sqrt{\frac{\pi}{2}} \text{erfc}
\left(\frac{\eta}{\sqrt{2}}\right). \label{App:MI_1}
\end{eqnarray}
which indeed is the solution of (\ref{App:governing_equation_Fn})
for $n=-1$. The sequence of steps listed in (\ref{App:MI_1}) may
in fact be considered a simple derivation of the well-known
Craig's formula \citep{craig1991new}, the first expression on the
righthand side of (\ref{App:MI_1}) being another form of the
Gaussian distribution function.

Next we show that following relationship holds between
$F_{n+1}(\eta)$ and $F_n(\eta)$
\begin{eqnarray}
F_{n+1}(\eta)&&=\int_0^{\pi}\frac{1}{\sqrt{2\pi}}\exp\left(-\frac{\eta^2}{2\cos^2\left(\frac{\alpha}{2}\right)}\right)
\sin^{2n+4}\left(\frac{\alpha}{2}\right)\mbox{d}\alpha \nonumber \\
&&=\int_\eta^{\infty}\left\{\int_{0}^{\pi}\frac{\eta_1}{\sqrt{2
\pi}\cos^2{\left(\frac{\alpha}{2}\right)}}
\exp\left(-\frac{\eta_1^2}{2\cos^2\left(\frac{\alpha}{2}\right)}\right)\sin^{2n+4}\left(\frac{\alpha}{2}\right)
\mbox{d}\alpha \right\}\mbox{d}\eta_1 \nonumber \\
&&=\int_\eta^{\infty}\left\{-\left.\text{erfc}
\left(\frac{\eta_1}{\sqrt2 \cos \left( \frac{\alpha}{2} \right) }
\right)
\sin^{2n+3} \left(\frac{\alpha}{2}\right)\right |_0^\pi + \right. \nonumber \\
&&\qquad \quad \left. \frac{2n+3}{2}\int_{0}^{\pi} \text{erfc}
\left(\frac{\eta_1}{\sqrt2 \cos \left( \frac{\alpha}{2} \right) }
\right)
\sin^{2n+2}\left(\frac{\alpha}{2}\right) \cos\left(\frac{\alpha}{2}\right) \mbox{d}\alpha \right\} \mbox{d}\eta_1 \nonumber \\
&&= \! (2n \!+ \!3) \! \int_\eta^\infty  \!\int_{\eta_1}^\infty \!
\left\{ \!  \int_0^{\pi}\frac{1}{\sqrt{2\pi}}\!
\exp\left(-\frac{\eta_2^2}{2\cos^2\left(\!\frac{\alpha}{2}\!
\right)}\right)\!\sin^{2n+2}\left(\!\frac{\alpha}{2}\!\right)
\mbox{d}\alpha \! \right\} \! \mbox{d}\eta_2\; \mbox{d}\eta_1 \nonumber \\
&& = (2n + 3)  \int_\eta^\infty  \int_{\eta_1}^\infty F_n(\eta_2)
\mbox{d}\eta_2 \; \mbox{d}\eta_1. \label{App:MI_relation}
\end{eqnarray}
Consequently
\begin{eqnarray}
F_{n+1}^{\prime}(\eta) = -(2n + 3) \int_{\eta}^\infty F_n(z)
\mbox{d}z \qquad \mbox{and} \quad F_{n+1}^{\prime\prime}(\eta) =
(2n + 3)F_n(\eta). \label{App:MI_relation-b}
\end{eqnarray}
Integrating the expression (\ref{App:governing_equation_Fn}) twice
we obtain
\begin{eqnarray}
F_n(\eta) - \eta \int_\eta^\infty F_n(\eta_1)
\mbox{d}\eta_1-\int_\eta^\infty \int_{\eta_1}^\infty
(2n+4)F_n(\eta_2) \mbox{d}\eta_2  \mbox{d}\eta_1=0.
\label{App:MI_relation_2}
\end{eqnarray}
Using relationships (\ref{App:MI_relation}),
(\ref{App:MI_relation-b}) and (\ref{App:MI_relation_2}) one finds
that if (\ref{App:Similarity_sol_Fn}) is the solution of
(\ref{App:governing_equation_Fn}) for some integer $n$ then the
result holds for $n+1$ as well. The proof therefore follows from
induction.

We next determine a solution to
(\ref{eqn:governing_equation_ustar}) that satisfies all the three
boundary conditions (\ref{eqn:boundary_conditions_ustar}a-c). We
begin with an evaluation of $F_n^\prime(0)$ for both $\beta$ and
$H_n$ depend on it:
\begin{eqnarray}
F_n^\prime(0)&&=\left.\int_0^{\pi}\frac{-\eta}{\sqrt{2\pi} \cos^2
\left(\frac{\alpha}{2}\right)}
\exp\left(-\frac{\eta^2}{2\cos^2\left(\frac{\alpha}{2}\right)}\right)\sin^{2n+2}\left(\frac{\alpha}{2}\right)
\mbox{d}\alpha \right|_{\eta = 0} \;. \label{App:Derivative_Fn_1}
\end{eqnarray}
Since the above integrand is singular at $\eta = 0$, we first
simplify (\ref{App:Derivative_Fn_1}) using integration by parts
and subsequently apply the limit $\eta = 0$ to the resulting
well-behaved integrand.
\begin{eqnarray}
F_n^\prime(0) && =\left. - \frac{2n+1}{2} \int_0^\pi \text{erfc} \left(\frac{\eta}{\sqrt2 \cos \left( \frac{\alpha}{2} \right) }  \right) \sin^{2n}\left(\frac{\alpha}{2}\right) \cos \left(\frac{\alpha}{2}\right) \mbox{d} \alpha\right|_{\eta = 0} \nonumber \\
&& = - \frac{2n+1}{2} \; \text{B}\left(\frac{2n+1}{2},1\right) \nonumber \\
&& = -1 \;, \label{App:Derivative_Fn_2}
\end{eqnarray}
here $\text{B}(x, y)$ denotes the Beta function.

Next we recast the boundary condition
(\ref{eqn:boundary_conditions_ustar}c) as follows
\begin{subequations}
\begin{eqnarray}
4 \sin \phi = \sum_{n=0}^\infty w_n
\left(\sin\frac{\phi}{2}\right)^{2n+1},
\label{App:superposition_1}
\end{eqnarray}
where
\begin{eqnarray}
 w_n = 2^{3-n}\prod_{i=1}^n \frac{(2i-3)}{i}.
 \label{App:superposition_1_weights}
\end{eqnarray}
\end{subequations}
Since the governing equation (\ref{eqn:governing_equation_ustar})
is linear, we may now superpose the similarity solutions
$\tilde{u}_{\phi, n}^\ast$ using the weights
(\ref{App:superposition_1_weights}) to obtain the solution of the
scaled azimuthal correction velocity component
($\tilde{u}_\phi^\ast$) as follows
\begin{eqnarray}
 \tilde{u}_\phi^\ast = \sum_{n=0}^{\infty} w_n \tilde{u}_{\phi, n}^\ast = \sum_{n=0}^{\infty} w_n H_n(\phi) F_n(\eta)
 \qquad \qquad \qquad \qquad \qquad\qquad \qquad \qquad && \nonumber \\
=\frac{-4}{\sqrt{\pi}}\tan\left(\frac{\phi}{2}\right)\int_0^{\pi}\exp\left(\frac{-\eta^2}{2\cos^2\left(\frac{\alpha}{2}\right)
}\right)\sin^2\left(\frac{\alpha}{2}\right)
\sqrt{1-\sin^2\left(\frac{\alpha}{2}\right)\sin^2\left(\frac{\phi}{2}\right)}\;
\mbox{d}\alpha, && \label{App:superposition_2}
\end{eqnarray}
where $\eta = \sqrt{2} y^\ast \cos\frac{\phi}{2}$.

\section{Numerical solution of the non-dimensional distortion flow streamfunction equation
(\ref{eqn:stag_dist_strm})}\label{sec:Appendix-B}

For computational convenience, we first apply the following
transformation to the governing equation
(\ref{eqn:stag_dist_strm}) for the non-dimensional distortion flow
streamfunction $\tilde{\psi}_s^\ast$:
\begin{equation}
\specialnumber{a,b} \sigma = \theta_s^\ast y_s^\ast, \qquad  \tau
= \frac{\theta_s^{\ast 2} - y_s^{\ast 2}}{2}.
\end{equation}
Rewritten in ($\sigma, \tau$) coordinates, equation
(\ref{eqn:stag_dist_strm}) assumes the form
\begin{eqnarray}
 \left(\frac{\partial^2 \tilde{\psi}_s^\ast}{\partial \sigma^2} +
 \frac{\partial^2 \tilde{\psi}_s^\ast}{\partial \tau^2}\right)  =
 \frac{1}{\sqrt{\pi}}\int_0^{\pi} \frac{\tilde{\psi}_s^\ast + 2\sigma}{(\sigma^2 + \tau^2)^{\frac{1}{2}}}
 \exp\left(\frac{-(\tilde{\psi}_s^\ast + 2\sigma)^2}{16 \cos^2\left(\frac{\alpha}{2}\right)}\right)
 \frac{\sin^2\left(\frac{\alpha}{2}\right)}{\cos\left(\frac{\alpha}{2}\right)}\mbox{d}\alpha,
\label{App:stag_dist_strm}
\end{eqnarray} while the boundary conditions
(\ref{eqn:boundary_conditions_stag}) transform as follows:
\begin{subequations}
\begin{eqnarray}
&&\tilde{\psi}_s^\ast = 0 \quad \text{at} \quad \sigma = 0, \\
&&\frac{\partial \tilde{\psi}_s^\ast}{\partial \sigma} = 0  \quad
\text{as} \quad \sigma \to \infty,
\end{eqnarray}
\begin{eqnarray}
&&\frac{\partial \tilde{\psi}_s^\ast}{\partial \tau} = 0  \quad \text{at} \quad \tau = 0, \\
&&\frac{\partial \tilde{\psi}_s^\ast}{\partial \tau} = 0  \quad
\text{as} \quad \tau \to \infty,
\end{eqnarray}
\label{App:boundary_conditions_stag}
\end{subequations} with the
boundary condition (\ref{App:boundary_conditions_stag}a-d)
implying impermeability, uniformity of the stream function outside
the thin stagnation layer, symmetry of the flow about the line
$\theta_s^\ast = y_s^\ast$ and unidirectionality of the flow
towards the end stage of the boundary layer, respectively.

To solve (\ref{App:stag_dist_strm}) we employ an iterative
process. We begin with a uniform initial guess state corresponding
to $\tilde{\psi}_s^\ast = 0$ everywhere. Subsequently, in each
iteration we use the present known $\tilde{\psi}_s^\ast$ to
compute the right hand side of (\ref{App:stag_dist_strm}). With
this known right hand side we solve the Poisson equation for
$\tilde{\psi}_s^\ast$ along with boundary conditions
(\ref{App:boundary_conditions_stag}a-d). We repeat this iterative
process until convergence is achieved (the maximum pointwise
difference in $\tilde{\psi}_s^\ast$ between successive iterations
is less than $10^{-10}$).

To discretize the Poisson equation $\tilde{\psi}_s^\ast$, we
employ standard second-order central finite differences on
exponentially stretched grids along both $\sigma$ and $\tau$
directions. Exponential stretching enables solution-adaptivity so
that the resulting spatial resolution is the highest in the
regions associated with large vorticity gradients. To solve for
the vector of unknowns consisting of discrete pointwise
$\tilde{\psi}_s^\ast$, we employ a specialized Gaussian
Elimination method that exploits the sparsity of the system to
reduce the overall computational expense.

To determine the tangential surface velocity at the shear-free
cylindrical boundary we evaluate the following relationship
\begin{eqnarray}
\tilde{u}_{\theta_{(s)}}^\ast = - \left.\frac{\partial
\tilde{\psi}_s^\ast}{\partial y_s^\ast}\right|_{y_s^\ast = 0} = -
\left.\sqrt{2 \tau} \frac{\partial \tilde{\psi}_s^\ast}{\partial
\sigma}\right|_{\sigma = 0}.
\end{eqnarray} using a sixth-order one-sided finite difference
approximation for the first derivative. The vorticity in the
stagnation region is conveniently computed from the converged
$\tilde{\psi}_s^\ast$ via the following transformed variant of the
relationship (\ref{eqn:stag_dist_strm}):
\begin{eqnarray}
\omega_s^\ast =
 -\frac{2}{\sqrt{\pi}}\int_0^{\pi} (\tilde{\psi}_s^\ast + 2\sigma)
 \exp\left(\frac{-(\tilde{\psi}_s^\ast + 2\sigma)^2}{16 \cos^2\left(\frac{\alpha}{2}\right)}\right)
 \frac{\sin^2\left(\frac{\alpha}{2}\right)}{\cos\left(\frac{\alpha}{2}\right)}\mbox{d}\alpha.
\end{eqnarray}

Likewise, in ($\sigma, \tau$) coordinates, the integral in the
expression (\ref{eqn:drag_dissipation_8}) assumes the following
form
\begin{eqnarray}
\int_{\theta_s^\ast = 0}^{\infty} \int_{y_s^\ast = 0}^{\infty} (
\omega_s^{\ast 2} - \omega_{bl}^{\ast 2}|_{\theta \to 0}) \;
\mbox{d}y_s^\ast \mbox{d}\;\theta_s^\ast = 2 \int_{\sigma =
0}^{\infty} \int_{\tau = 0}^{\infty} \frac{( \omega_s^{\ast 2} -
\omega_{bl}^{\ast 2}|_{\theta \to 0})}{2(\sigma^2+\tau^2)^{1/2}}
\; \mbox{d}\sigma \; \mbox{d}\tau.
\end{eqnarray} To evaluate this integral numerically we apply a two-dimensional
generalization of the Simpson's quadrature rule.

\section{Derivation of the estimate (\ref{eqn:drag_dissipation_7})}
\label{sec:Appendix-C} To establish the result
(\ref{eqn:drag_dissipation_7}) we first consider evaluation of the
leading order term in the following integral
\begin{eqnarray}
\text{I}(A, n, s_f, \delta, M(s), N(s)) = \int_{s=0}^{s_f} s^2
M(s) \int_{t=0}^{\frac{A s^n}{\delta}} t^2 \exp \left(-s^2 t^2
N^2(s) \right)  \mbox{d}t \; \mbox{d}s, \label{App:Int_1}
\end{eqnarray}
in the limit of $\delta \to 0^+$, where $n>0, \quad A > 0, \quad
N(s) > 0 \; \forall \;s \; \in[0,s_f]$. Subsequently we show that
(\ref{eqn:drag_dissipation_7}) is a special case of the general
integral (\ref{App:Int_1}).

We denote the limiting expressions
\begin{equation}
M(s) \to M_0, \; \frac{\mbox{d} M(s)}{\mbox{d} s} \to M^\prime_0,
\; N(s) \to N_0, \text{ and } \frac{\mbox{d} N(s)}{\mbox{d} s} \to
N^\prime_0 \quad \text{as} \quad s \to 0, \;
\label{App:Int_1_asmp}
\end{equation}
where $M_0$, $N_0$, $M^\prime_0$  and $N^\prime_0$ are all
bounded. Performing the integration with respect to $t$ in
(\ref{App:Int_1}) yields
\begin{eqnarray}
\text{I} \! = \! \int_{s=0}^{s_f} \!  \frac{A M(s)}{4sN^2(s)} \!
\left[  \!  \frac{-2 s^{n+1}}{\delta} \! \exp \left( \!
\frac{-s^{2n+2} A^2 N^2(s)}{\delta^2} \!  \right) \! +\!
\frac{\sqrt{\pi}}{A N(s)}\text{erf} \left( \!  \frac{s^{n+1} A
N(s)}{\delta} \!  \right) \! \right] \! \mbox{d} \! s.
\label{App:Int_2}
\end{eqnarray}
We split the above integral into two parts with limits of
integration in the first and second parts ranging from $s=0$ to
$s=\epsilon$ and $s=\epsilon$ to $s=s_f$, respectively. We set
$\epsilon$ such that $\delta^{\frac{1}{n+1}} \ll \epsilon $ and
$\epsilon \ll 1$. This choice ensures that the integrand of the
first part can be simplified by using the limit $\epsilon \to 0$,
while the integrand in the second part can be simplified by taking
the limit $\epsilon^{n+1} / \delta \to \infty$.

Simplification of the first part proceeds as follows
\begin{eqnarray}
\text{I}_1 \! &&= \! \int_{s=0}^{\epsilon} \!  \frac{A
M(s)}{4sN^2(s)} \!  \left[  \!  \frac{-2 s^{n+1}}{\delta} \! \exp
\left( \!  \frac{-s^{2n+2} A^2 N^2(s)}{\delta^2} \!  \right) \!
+\! \frac{\sqrt{\pi}}{A N(s)}\text{erf}
\left( \!  \frac{s^{n+1} A N(s)}{\delta} \!  \right) \! \right] \! \mbox{d} \! s   \nonumber \\
&&=\int_{s=0}^{\epsilon} \frac{A M_0}{4sN^2_0}\left[  -2
\frac{s^{n+1}}{\delta} \exp \left(-A^2 N^2_0
\frac{s^{2n+2}}{\delta^2} \right)  + \frac{\sqrt{\pi}}{A
N_0}\text{erf} \left(A N_0 \frac{s^{n+1}}{\delta}\right)  \right]
\mbox{d}s
\nonumber \\
 && \qquad \qquad \qquad \qquad \qquad \qquad \qquad \qquad \qquad \qquad \qquad \qquad+ \mbox{O}\left(\epsilon \ln
 \left[\frac{\epsilon^{n+1}}{\delta}\right]\right)  \nonumber \\
&& \approx \frac{-\sqrt{\pi} M_0}{4 (n+1) N_0^3} \text{erf}
\left(A N_0 \frac{\epsilon^{n+1}}{\delta}\right) +
\frac{\sqrt{\pi} M_0}{4 (n+1) N_0^3} \int_0^{\frac{A N_0
\epsilon^{n+1}}{\delta}} \frac{\text{erf} \; (\xi) }{\xi} d\xi.
\label{App:I1_1}
\end{eqnarray}
From (\ref{App:MI_1}) we have the following relation
\begin{eqnarray}
\text{erf}(\xi) = 1 -
\frac{1}{\pi}\int_0^{\pi}\exp\left(-\frac{\xi^2}{\cos^2\frac{\alpha}{2}}\right)\mbox{d}\alpha.
\end{eqnarray}
Using this relation in (\ref{App:I1_1}) and performing an
integration with respect to $\xi$ we obtain
\begin{eqnarray}
\text{I}_1 \approx \frac{\sqrt{\pi} M_0}{4 (n+1) N_0^3} \! \left(
\! - \text{erf} \left(A N_0 \frac{\epsilon^{n+1}}{\delta}\right)
\! +  \!  \left[ \ln \xi +  \frac{1}{2 \pi} \int_0^\pi \text{E}_1
\left(\frac{\xi^2}{\cos^2 \frac{\alpha}{2}}\right) \mbox{d} \alpha
\right]_{\xi = 0}^{\frac{A N_0 \epsilon^{n+1}}{\delta}} \right),
\label{App:I1_2}
\end{eqnarray}
where $\text{E}_1$ denotes the exponential integral function.
Using $\text{E}_1(x) \sim - \gamma - \ln x$ as $x \to 0$, where
$\gamma = 0.577$ is the Euler's constant
\citep{abramowitz1964handbook}, we simplify the limits in
(\ref{App:I1_2}) and obtain
\begin{eqnarray}
\text{I}_1 \approx \frac{\sqrt{\pi} M_0}{4 (n+1) N_0^3} \left( -
\text{erf} \left(A N_0 \frac{\epsilon^{n+1}}{\delta}\right) +
\left[\frac{\gamma}{2} +  \ln \left(\frac{2 A N_0
\epsilon^{n+1}}{\delta}\right) \right. \right.
&&  \nonumber \\
\left. \left. +  \frac{1}{2 \pi} \int_0^\pi \text{E}_1
\left(\frac{A^2 N_0^2 \epsilon^{2n+2}}{\delta^2 \cos^2
\frac{\alpha}{2}}\right) d \alpha  \right] \right). &&
\label{App:I1_3}
\end{eqnarray}
Using the fact that $\epsilon^{n+1} \gg \delta $, $\text{erf}(x)
\sim 1 - \exp(-x^2)/\sqrt{\pi} x$ and $\text{E}_1(x) \sim \exp(-x)
/ x$ as $x \to \infty$, we get the leading order terms in
(\ref{App:I1_3}) as follows:
\begin{eqnarray}
\text{I}_1 = \frac{\sqrt{\pi} M_0}{4 (n+1) N_0^3} \left(-1 +
\frac{\gamma}{2} + \ln \left(\frac{2 A N_0
\epsilon^{n+1}}{\delta}\right) \right)
 + \mbox{O}\left(\frac{\exp(-A^2 N_0^2 \delta^{-2} \epsilon^{2n+2})}{\delta^{-1} \epsilon^{n+1}}\right) &&
 \nonumber \\ +  \mbox{O}\left(\epsilon \ln \left(\frac{\epsilon^{n+1}}{\delta}\right)\right).
 \label{App:I1_4}
\end{eqnarray}

Next we simplify the second part $\text{I}_2$
\begin{eqnarray}
\text{I}_2 \! = \! \int_{s=\epsilon}^{s_f} \!  \frac{A
M(s)}{4sN^2(s)} \!  \left(  \!  \frac{-2 s^{n+1}}{\delta} \!  \exp
\left( \!  \frac{-s^{2n+2} A^2 N^2(s)}{\delta^2} \!  \right) \!
+\! \frac{\sqrt{\pi}}{A N(s)} \text{erf} \left( \!  \frac{s^{n+1}
A N(s)}{\delta} \!  \right) \! \right) \! \mbox{d} \! s
\label{App:I2_1}
\end{eqnarray}
Using $s^{n+1} / \delta \gg 1$ we expand the integrand in
(\ref{App:I2_1}) as follows
\begin{eqnarray}
\text{I}_2 && =  \int_{s=\epsilon}^{s_f} \frac{A
M(s)}{4sN^2(s)}\left( \frac{\sqrt{\pi}}{A N(s)} -2
\frac{s^{n+1}}{\delta} \exp \left(-A^2 N^2(s) \frac{s^{2n+2}}{\delta^2} \right) \dots \right) \mbox{d} s \nonumber \\
&& = \int_{s=\epsilon}^{s_f} \frac{\sqrt{\pi} M(s)}{4sN^3(s)}
\mbox{d}s + \mbox{O}\left(\frac{\epsilon^{n+1}
\exp(-A^2 N_0^2 \delta^{-2} \epsilon^{2n+2})}{\delta}\right) \nonumber \\
&& \approx \int_{s=\epsilon}^{s_f} \frac{\sqrt{\pi}
M(s)}{4sN^3(s)}  \mbox{d}s . \label{App:I2_2}
\end{eqnarray}
Next we apply a singularity subtraction technique to regularize
the integrand in (\ref{App:I2_2}) as follows
\begin{eqnarray}
\text{I}_2 \approx \int_{s=s_0}^{s=s_f}\frac{\sqrt{\pi} M(s)}{4 s N^3(s)} \mbox{d}s +\int_{s = \epsilon}^{s_0} \frac{\sqrt{\pi}}{4 s}
\left(\frac{M(s)}{N^3(s)} - \frac{M_0}{N^3_0}\right) \mbox{d}s + \int_{s=\epsilon}^{s_0} \frac{\sqrt{\pi} M_0}{4sN^3_0}  \mbox{d}s, && \nonumber \\
\label{App:I2_3}
\end{eqnarray}
where $s_0 \leqslant s_f$ is an $O(1)$ limit that can be set more
or less arbitrarily. Since the integrand in second part remains
bounded in the limit $s \to 0$, the leading order terms in
$\text{I}_2$ are readily found:
\begin{eqnarray}
\text{I}_2 = \int_{s=s_0}^{s=s_f}\frac{\sqrt{\pi} M(s)}{4 s
N^3(s)} \mbox{d}s +\int_{s = 0}^{s_0} \frac{\sqrt{\pi}}{4 s}
\left(\frac{M(s)}{N^3(s)} - \frac{M_0}{N^3_0}\right) \mbox{d}s +
\frac{\sqrt{\pi} M_0}{4 N_0^3}
\ln \left(\frac{s_0}{\epsilon}\right) && \nonumber \\
+ \mbox{O}\left(\frac{  \epsilon^{n+1} \exp(-A^2 N_0^2 \delta^{-2}
\epsilon^{2n+2})}{\delta}\right) +  \mbox{O}\left(\epsilon\right).
\label{App:I2_4}
\end{eqnarray}
Adding the results (\ref{App:I1_3}) and (\ref{App:I2_4}) we deduce
the leading order terms in (\ref{App:Int_1}) as follows
\begin{eqnarray}
\text{I} \approx \int_{s=s_0}^{s=s_f}\frac{\sqrt{\pi} M(s)}{4 s
N^3(s)} \mbox{d}s +
\int_{s = 0}^{s_0} \frac{\sqrt{\pi}}{4 s}   \left(\frac{M(s)}{N^3(s)} - \frac{M_0}{N^3_0}\right) \mbox{d}s && \nonumber \\
+ \frac{\sqrt{\pi} M_0}{4 (n+1) N_0^3} \left(\frac{\gamma}{2} - 1
+ \ln \left(\frac{2 A N_0 s_0^{n+1}}{\delta}\right) \right).&&
\label{App:Int_3}
\end{eqnarray}
Thus, our analysis of the leading order terms separates out the
$\delta$-dependence of the integral $I$. The remaining integrals
in (\ref{App:Int_3}) involve non-singular integrands and are
therefore amenable to accurate approximation using standard
quadrature rules.

Next we show that the integral (\ref{eqn:drag_dissipation_7}) is
indeed a special case of the general form (\ref{App:Int_1}) and
can therefore be estimated in a manner similar to the one outlined
in the foregoing analysis. Expressed in the $\theta$ coordinate,
the square of vorticity from the boundary layer region assumes the
following form
\begin{eqnarray}
\omega_{bl}^{\ast 2} =\frac{64}{\pi}  \int_{0}^{\pi}
\int_{0}^{\pi} y^{\ast  2} \sin^2 \left(\frac{\theta}{2}\right)
\cos^2 \left(\frac{\theta}{2}\right) \exp\left(- y^{\ast 2}
\left[\frac{\sin^2
\left(\frac{\theta}{2}\right)}{\cos^2\left(\frac{\alpha_1}{2}\right)}
+ \frac{ \sin^2
\left(\frac{\theta}{2}\right)}{\cos^2\left(\frac{\alpha_2}{2}\right)}\right]
\right)
\times \qquad &&   \nonumber \\
  \tan^2\left(  \frac{\alpha_1}{2} \right)  \tan^2\left(  \frac{\alpha_2}{2}  \right)  \sqrt{1  -
  \sin^2\left(  \frac{\alpha_1}{2} \right) \cos^2\left(\frac{\theta}{2}\right)}
\sqrt{1  -  \sin^2\left(  \frac{\alpha_2}{2}  \right)
\cos^2\left(\frac{\theta}{2}\right)} \; \mbox{d}\alpha_1 \;
\mbox{d}\alpha_2 \qquad && \label{App:Omega_bl^2}
\end{eqnarray}
From the above relationship, we can recast the boundary layer
contribution to the vorticity-squared integral in
(\ref{eqn:drag_dissipation_7}) as follows
\begin{eqnarray}
\int_{\theta = 0}^{\pi} \int_{y^\ast =
0}^{\frac{1}{\delta}(\frac{\theta}{A})^{1/b}} \omega_{bl}^{\ast 2}
\; \mbox{d}y^\ast \mbox{d}\theta =
\qquad \qquad \qquad \qquad \qquad \qquad \qquad \qquad \qquad \quad  &&\nonumber \\
\frac{64}{\pi} \int_0^\pi \int_0^\pi \tan^2\left(
\frac{\alpha_1}{2} \right) \tan^2\left(  \frac{\alpha_2}{2}
\right) \; \text{I}(1/A^{1/b}, 1/b, \pi, \delta, M(\theta),
N(\theta)) \; \mbox{d} \alpha_1 \mbox{d} \alpha_2, &&
\label{App:Omega_bl^2_int}
\end{eqnarray}
where
\begin{eqnarray}
&& M(\theta) = \frac{\sin^2
\left(\frac{\theta}{2}\right)}{\theta^2} \cos^2
\left(\frac{\theta}{2}\right) \sqrt{1  -  \sin^2\left(
\frac{\alpha_1}{2} \!\right) \cos^2\left(\frac{\theta}{2}\right)}
\!
\sqrt{1 \! - \! \sin^2\left( \! \frac{\alpha_2}{2} \! \right) \cos^2\left(\frac{\theta}{2}\right)} \nonumber \\
&& \text{and} \qquad N(\theta) = \frac{\sin
\left(\frac{\theta}{2}\right)}{\theta}
\sqrt{\frac{1}{\cos^2\left(\frac{\alpha_1}{2}\right)} + \frac{
1}{\cos^2\left(\frac{\alpha_2}{2}\right)}}.
\label{App:Omega_bl^2_int_cmp}
\end{eqnarray}
Likewise, the wake contribution to the vorticity-squared integral
in (\ref{eqn:drag_dissipation_7}) can be expressed in the general
form (\ref{App:Int_1}), as shown below
\begin{eqnarray}
\int_{x = 0}^{\infty} \int_{z^\ast = 0}^{\frac{A x^{b}}{\delta}}
\omega_{w}^{\ast 2} \; \mbox{d}z^\ast \mbox{d}x =
\qquad \qquad \qquad \qquad \qquad \qquad \qquad \qquad && \nonumber \\
\frac{2}{\pi} \int_{-2}^{2} \int_{-2}^{2} \sqrt{4 - \kappa_1^2}
\sqrt{4 - \kappa_2^2} \; \text{I}(A, b, \infty, \delta, M(x),
N(x)) \; \mbox{d}\kappa_1 \mbox{d}\kappa_2,&&
\label{App:Omega_w^2_int}
\end{eqnarray}
where
\begin{eqnarray}
&& M(x) = \frac{\left(1 - \frac{1}{(x+ 1)^2}\right)^2}{2 x^2
\left( x+ 1 + \frac{1}{x+ 1} - \kappa_1\right)^{3/2} \left( x+ 1 + \frac{1}{x+ 1} - \kappa_2\right)^{3/2}} \nonumber \\
&& \text{and} \qquad N(x) = \frac{\left(1 - \frac{1}{(x+
1)^2}\right)}{2 x} \sqrt{\frac{1}{ x+ 1 + \frac{1}{x+ 1} -
\kappa_1} + \frac{1}{ x+ 1 + \frac{1}{x+ 1} - \kappa_2}}.
\label{App:Omega_w^2_int_cmp}
\end{eqnarray}

Making use of the foregoing analysis in the derivation of the
leading order terms and in the quadrature-based numerical
approximation of the remaining regularized integrals in
(\ref{App:Omega_bl^2_int}) and (\ref{App:Omega_w^2_int}), we
readily establish the estimate given by
(\ref{eqn:drag_dissipation_7}).

\UseRawInputEncoding
\bibliographystyle{jfm}
\bibliography{Anuj-Cyl-Shear-Free}

\begin{thebibliography}{41}
\expandafter\ifx\csname natexlab\endcsname\relax\def\natexlab#1{#1}\fi
\def\au#1{#1} \def\ed#1{#1} \def\yr#1{#1}\def\at#1{#1}\def\jt#1{\textit{#1}}
  \def\bt#1{#1}\def\bvol#1{\textbf{#1}} \def\vol#1{#1} \def\pg#1{#1}
  \def\publ#1{#1}\def\arxiv#1{#1}\def\org#1{#1}\def\st#1{\textit{#1}}

\bibitem[Abramowitz \& Stegun(1968)]{abramowitz1964handbook}
{\sc \au{Abramowitz, M.} \& \au{Stegun, I.~A.}} \yr{1968} {\em Handbook of
  mathematical functions: with formulas, graphs, and mathematical tables\/}.
  \publ{Dover}.

\bibitem[Arakeri \& Shukla(2013)]{arakeri2013unified}
{\sc \au{Arakeri, J.~H.} \& \au{Shukla, R.~K.}} \yr{2013}  \at{A unified view
  of energetic efficiency in active drag reduction, thrust generation and
  self-propulsion through a loss coefficient with some applications}.  \jt{J.
  Fluid. Struct.}  \bvol{41},  \pg{22--32}.

\bibitem[Batchelor(2000)]{batchelor2000introduction}
{\sc \au{Batchelor, G.~K.}} \yr{2000} {\em An introduction to fluid
  dynamics\/}.  \publ{Cambridge University Press}.

\bibitem[Bocquet \& Lauga(2011)]{lauga2011natmat}
{\sc \au{Bocquet, L.} \& \au{Lauga, E.}} \yr{2011}  \at{A smooth future?}
  \jt{Nat. Mater.}  \bvol{10},  \pg{334--337}.

\bibitem[Craig(1991)]{craig1991new}
{\sc \au{Craig, J.~W.}} \yr{1991} A new, simple and exact result for
  calculating the probability of error for two-dimensional signal
  constellations.  \bt{In {\em Military Communications Conference, 1991.
  MILCOM'91, Conference Record, Military Communications in a Changing World.,
  IEEE\/}},  \pg{pp. 571--575}. IEEE.

\bibitem[Haase {\em et~al.\/}(2015)Haase, Chapman, Tsai, Lohse \&
  Lammertink]{Lohse2015slip}
{\sc \au{Haase, S.~A.}, \au{Chapman, S.~J.}, \au{Tsai, P.~A.}, \au{Lohse, D.}
  \& \au{Lammertink, R. G.~H.}} \yr{2015}  \at{The {G}raetz--{N}usselt problem
  extended to continuum flows with finite slip}.  \jt{J. Fluid Mech.}
  \bvol{764},  \pg{R3}.

\bibitem[Haase \& Lammertink(2016)]{Haase2015slip}
{\sc \au{Haase, S.~A.} \& \au{Lammertink, R. G.~H.}} \yr{2016}  \at{Heat and
  mass transfer over slippery, superhydrophobic surfaces}.  \jt{Phys. Fluids}
  \bvol{28},  \pg{042002}.

\bibitem[Harper(1963)]{harper1963boundary}
{\sc \au{Harper, J.~F.}} \yr{1963}  \at{On boundary layers in two-dimensional
  flow with vorticity}.  \jt{J. Fluid Mech.}  \bvol{17}~(1),  \pg{141--153}.

\bibitem[Harper(1972)]{harper1972review}
{\sc \au{Harper, J.~F.}} \yr{1972}  \at{The motion of bubbles and drops through
  liquids}.  \jt{Adv. Appl. Mech.}  \bvol{12},  \pg{59--129}.

\bibitem[Harper \& Moore(1968)]{harper1968motion}
{\sc \au{Harper, J.~F.} \& \au{Moore, D.~W.}} \yr{1968}  \at{The motion of a
  spherical liquid drop at high {R}eynolds number}.  \jt{J. Fluid Mech.}
  \bvol{32}~(2),  \pg{367--391}.

\bibitem[Hinch(1991)]{Hinch1991Book}
{\sc \au{Hinch, E.~J.}} \yr{1991} {\em Perturbation methods\/}.
  \publ{Cambridge University Press}.

\bibitem[Hugues \& Randriamampianina(1998)]{hugues1998improved}
{\sc \au{Hugues, S.} \& \au{Randriamampianina, A.}} \yr{1998}  \at{An improved
  projection scheme applied to pseudospectral methods for the incompressible
  navier--stokes equations}.  \jt{Intl J Numer. Meth. Fluids}  \bvol{28}~(3),
  \pg{501--521}.

\bibitem[Joseph {\em et~al.\/}(2007)Joseph, Funada \& Wang]{Joseph2007Book}
{\sc \au{Joseph, D.~D.}, \au{Funada, T.} \& \au{Wang, J.}} \yr{2007} {\em
  Potential flows of viscous and viscoelastic fluids\/}.  \publ{Cambridge
  University Press}.

\bibitem[Kang \& Leal(1988)]{kang1988drag}
{\sc \au{Kang, I.~S.} \& \au{Leal, L.~G.}} \yr{1988}  \at{The drag coefficient
  for a spherical bubble in a uniform streaming flow}.  \jt{Phys. Fluids}
  \bvol{31}~(2),  \pg{233--237}.

\bibitem[Karatay {\em et~al.\/}(2013)Karatay, Haase, Visser, Sun, Lohse, Tsai
  \& Lammertink]{lohse2013slip}
{\sc \au{Karatay, E.}, \au{Haase, A.~S.}, \au{Visser, C.~W.}, \au{Sun, C.},
  \au{Lohse, D.}, \au{Tsai, P.~A.} \& \au{Lammertink, R. G.~H.}} \yr{2013}
  \at{Control of slippage with tunable bubble mattresses}.  \jt{Proc. Natl.
  Acad. Sci.}  \bvol{110}~(21),  \pg{8422--8426}.

\bibitem[von K\'{a}rm\'{a}n(1911)]{von1911mechanismus}
{\sc \au{von K\'{a}rm\'{a}n, T.}} \yr{1911}  \at{{\"U}ber den mechanismus des
  widerstandes, den ein bewegter k{\"o}rper in einer fl{\"u}ssigkeit
  erf{\"a}hrt}.  \jt{Nachrichten von der Gesellschaft der Wissenschaften zu
  G{\"o}ttingen, Mathematisch-Physikalische Klasse}  \bvol{1911},
  \pg{509--517}.

\bibitem[Lamb(1932)]{Lamb1932Book}
{\sc \au{Lamb, H.}} \yr{1932} {\em Hydrodynamics\/}.  \publ{Cambridge
  University Press}.

\bibitem[Leal(1989)]{leal1989vorticity}
{\sc \au{Leal, L.~G.}} \yr{1989}  \at{Vorticity transport and wake structure
  for bluff bodies at finite {R}eynolds number}.  \jt{Phys. Fluids}
  \bvol{1}~(1),  \pg{124--131}.

\bibitem[Leal(2007)]{leal2007advanced}
{\sc \au{Leal, L.~G.}} \yr{2007} {\em Advanced transport phenomena: fluid
  mechanics and convective transport processes\/}.  \publ{Cambridge University
  Press}.

\bibitem[Legendre {\em et~al.\/}(2009)Legendre, Lauga \&
  Magnaudet]{legendre2009influence}
{\sc \au{Legendre, D.}, \au{Lauga, E.} \& \au{Magnaudet, J.}} \yr{2009}
  \at{Influence of slip on the dynamics of two-dimensional wakes}.  \jt{J.
  Fluid Mech.}  \bvol{633},  \pg{437--447}.

\bibitem[Levich(1949)]{levich1949motion}
{\sc \au{Levich, V.~G.}} \yr{1949}  \at{The motion of bubbles at high
  {R}eynolds numbers}.  \jt{Zh. Eksp. Teor. Fiz}  \bvol{19}~(18),  \pg{436f}.

\bibitem[Li {\em et~al.\/}(2014)Li, Li, Xue, Yang, Su, Xia, Shi, Lin \&
  Duan]{li2014effect}
{\sc \au{Li, D.}, \au{Li, S.}, \au{Xue, Y.}, \au{Yang, Y.}, \au{Su, W.},
  \au{Xia, Z.}, \au{Shi, Y.}, \au{Lin, H.} \& \au{Duan, H.}} \yr{2014}  \at{The
  effect of slip distribution on flow past a circular cylinder}.  \jt{J. Fluid.
  Struct.}  \bvol{51},  \pg{211--224}.

\bibitem[Magnaudet \& Eames(2000)]{Magnaudet2000review}
{\sc \au{Magnaudet, J.} \& \au{Eames, I.}} \yr{2000}  \at{The motion of
  high-{R}eynolds-number bubbles in inhomogeneous flows}.  \jt{Annu. Rev. Fluid
  Mech.}  \bvol{32},  \pg{659--708}.

\bibitem[Michaelides(2003)]{Michaelides2003review}
{\sc \au{Michaelides, E.~E.}} \yr{2003}  \at{Hydrodynamic force and heat/mass
  transfer from particles, bubbles, and drops—-{T}he {F}reeman scholar
  lecture}.  \jt{ASME J. Fluids Eng.}  \bvol{125},  \pg{209--238}.

\bibitem[Moore(1963)]{JFM_Moore_1962}
{\sc \au{Moore, D.~W.}} \yr{1963}  \at{The boundary layer on a spherical gas
  bubble}.  \jt{J. Fluid Mech.}  \bvol{16}~(2),  \pg{161--176}.

\bibitem[Moore(1965)]{moore1965velocity}
{\sc \au{Moore, D.~W.}} \yr{1965}  \at{The velocity of rise of distorted gas
  bubbles in a liquid of small viscosity}.  \jt{J. Fluid Mech.}  \bvol{23}~(4),
   \pg{749--766}.

\bibitem[Muralidhar {\em et~al.\/}(2011)Muralidhar, Ferrer, Daniello \&
  Rothstein]{muralidhar2011influence}
{\sc \au{Muralidhar, P.}, \au{Ferrer, N.}, \au{Daniello, R.} \& \au{Rothstein,
  J.~P.}} \yr{2011}  \at{Influence of slip on the flow past superhydrophobic
  circular cylinders}.  \jt{J. Fluid Mech.}  \bvol{680},  \pg{459--476}.

\bibitem[Ou {\em et~al.\/}(2004)Ou, Perot \& Rothstein]{ou2004laminar}
{\sc \au{Ou, J.}, \au{Perot, B.} \& \au{Rothstein, J.~P.}} \yr{2004}
  \at{Laminar drag reduction in microchannels using ultrahydrophobic surfaces}.
   \jt{Phys. Fluids}  \bvol{16}~(12),  \pg{4635--4643}.

\bibitem[Rehman {\em et~al.\/}(2017)Rehman, Kumar \& Shukla]{shukla2017slip}
{\sc \au{Rehman, N. M.~A.}, \au{Kumar, A.} \& \au{Shukla, R.~K}} \yr{2017}
  \at{Influence of hydrodynamic slip on convective transport in flow past a
  circular cylinder}.  \jt{Theor. Comput. Fluid Dyn.}  \bvol{31},
  \pg{251--280}.

\bibitem[Rothstein(2010)]{rothstein2010slip}
{\sc \au{Rothstein, J.~P.}} \yr{2010}  \at{Slip on superhydrophobic surfaces}.
  \jt{Annu. Rev. Fluid Mech.}  \bvol{42},  \pg{89--109}.

\bibitem[Schlichting \& Gersten(2003)]{schlichting2003boundary}
{\sc \au{Schlichting, H.} \& \au{Gersten, K.}} \yr{2003} {\em Boundary Layer
  theory\/}.  \publ{Springer}.

\bibitem[Seo \& Song(2012)]{seo2012numerical}
{\sc \au{Seo, I.~W.} \& \au{Song, C.~G.}} \yr{2012}  \at{Numerical simulation
  of laminar flow past a circular cylinder with slip conditions}.  \jt{Int. J
  Numer. Meth. Fluids}  \bvol{68}~(12),  \pg{1538--1560}.

\bibitem[Shukla \& Arakeri(2013)]{shukla2013minimum}
{\sc \au{Shukla, R.~K.} \& \au{Arakeri, J.~H.}} \yr{2013}  \at{Minimum power
  consumption for drag reduction on a circular cylinder by tangential surface
  motion}.  \jt{J. Fluid Mech.}  \bvol{715},  \pg{597--641}.

\bibitem[Shukla {\em et~al.\/}(2007)Shukla, Tatineni \& Zhong]{shukla2007very}
{\sc \au{Shukla, R.~K.}, \au{Tatineni, M.} \& \au{Zhong, X.}} \yr{2007}
  \at{Very high-order compact finite difference schemes on non-uniform grids
  for incompressible navier--stokes equations}.  \jt{J. Comput. Phys.}
  \bvol{224}~(2),  \pg{1064--1094}.

\bibitem[Shukla \& Zhong(2005)]{shukla2005derivation}
{\sc \au{Shukla, R.~K} \& \au{Zhong, X.}} \yr{2005}  \at{Derivation of
  high-order compact finite difference schemes for non-uniform grid using
  polynomial interpolation}.  \jt{J. Comput. Phys.}  \bvol{204}~(2),
  \pg{404--429}.

\bibitem[Sooraj {\em et~al.\/}(2020)Sooraj, Ramagya, Khan, Sharma \&
  Agrawal]{Atul2020slip}
{\sc \au{Sooraj, P.}, \au{Ramagya, M.~S.}, \au{Khan, M.~H.}, \au{Sharma, A.} \&
  \au{Agrawal, A.}} \yr{2020}  \at{Effect of superhydrophobicity on the flow
  past a circular cylinder in various flow regimes}.  \jt{J. Fluid Mech.}
  \bvol{897},  \pg{A21}.

\bibitem[Strouhal(1878)]{strouhal1878ueber}
{\sc \au{Strouhal, V.}} \yr{1878} {\em Ueber eine besondere Art der
  Tonerregung\/}.

\bibitem[Van~Dyke(1975)]{van1975perturbation}
{\sc \au{Van~Dyke, M.}} \yr{1975} {\em Perturbation methods in fluid
  mechanics\/}.  \publ{The Parabolic Press}.

\bibitem[Williamson(1996)]{williamson1996vortex}
{\sc \au{Williamson, C. H.~K.}} \yr{1996}  \at{Vortex dynamics in the cylinder
  wake}.  \jt{Annu. Rev. Fluid Mech.}  \bvol{28},  \pg{477--539}.

\bibitem[Xiong \& Yang(2017)]{elongatedCylin2017slip}
{\sc \au{Xiong, Y.~L.} \& \au{Yang, D.}} \yr{2017}  \at{Influence of slip on
  the three-dimensional instability of flow past an elongated superhydrophobic
  bluff body}.  \jt{J. Fluid Mech.}  \bvol{814},  \pg{69--94}.

\bibitem[You \& Moin(2007)]{you2007effects}
{\sc \au{You, D.} \& \au{Moin, P.}} \yr{2007}  \at{Effects of hydrophobic
  surfaces on the drag and lift of a circular cylinder}.  \jt{Phys. Fluids}
  \bvol{19}~(8),  \pg{081701}.

\end{thebibliography}

\end{document}